%% file: main.tex
\documentclass{aa}

\usepackage{graphicx}
\usepackage{txfonts}
\usepackage{lipsum}
\usepackage{subcaption}         
\usepackage{lscape}             
\usepackage{placeins}           

\linenumbers
\renewcommand\makeLineNumber{}

\usepackage[hidelinks]{hyperref}

\usepackage{mathtools}

\usepackage{tikz}
\usetikzlibrary{decorations.pathmorphing}
\tikzset{
  feynman/.cd,
  every diagram/.style={x=+1.3em, y=+1.3em, baseline=+-.5ex},
  label/.style={anchor=north west,inner sep=+.4em, scale=.707,
    node font=\everymath\expandafter{\the\everymath\scriptstyle}},
  labelleft/.style={anchor=east,inner sep=+.4em, scale=.707,
    node font=\everymath\expandafter{\the\everymath\scriptstyle}},
  labelright/.style={anchor=west,inner sep=+.4em, scale=.707,
    node font=\everymath\expandafter{\the\everymath\scriptstyle}},
  labelupper/.style={anchor=south,inner sep=+.4em, scale=.707,
    node font=\everymath\expandafter{\the\everymath\scriptstyle}},
  labellower/.style={anchor=north,inner sep=+.4em, scale=.707,
    node font=\everymath\expandafter{\the\everymath\scriptstyle}},
  node/.style={shape=circle, inner sep=+0pt, minimum size=+2pt, node contents=},
  node ./.style={feynman/node, shape=coordinate},
  node o/.style={feynman/node, fill=white, draw},
  node */.style={feynman/node, fill,       draw},
  node +/.style={
    feynman/node, fill, draw, shape=rectangle, rotate=45, minimum size=+1pt},
  loop/.style n args={4}{insert path={
      (0,0) node[feynman/label]{$#4$} node[feynman/node #1] -- ++(right:1)
      arc[start angle=-90, delta angle=360, radius=.5] node[feynman/node #2]
      -- ++(right:1) node[feynman/node #3]}},
  cross diagram/.style={
    /tikz/feynman/label/.append style={anchor=south east, inner sep=+.2pt}},
  cross/.style n args={6}{insert path={
    (225:.707) node[feynman/label]{$#6$} node[feynman/node #4]
    -- (45:.707) node[midway,feynman/node #3] node[feynman/node #2]
    (135:.707) node[feynman/node #1] -- (-45:.707) node[feynman/node #5]}},
  line/.style n args={3}{insert path={
    (0,0) node[feynman/label]{$#3$} node[feynman/node #1]
    -- (right:.8) node[feynman/node #2]}}}
\newcommand*\tfInt[5]{%
  \text{\tikz[feynman/every diagram,feynman/#1 diagram/.try,#2]
    \draw[feynman/#1={#3{#4}{#5}}];}}

\tikzset{
  feynman/.cd,
  wedgeupper/.style n args={6}{insert path={
      (0, .5) node[feynman/labelleft]{$a$} node[feynman/node #1]
      (0,-.5) node[feynman/labelleft]{$b$} node[feynman/node #2]
      (1.8, .5) node[feynman/labelright]{$1$} node[feynman/node #3]
      (1.8,-.5) node[feynman/labelright]{$2$} node[feynman/node #4]
      (0.9, .5) node[feynman/labelupper]{$#5$}
      (0.9,-.1) node[feynman/labellower]{$#6$}
      (0, .5) -- (1.9, .5)
      (0, .5) -- (1.9, -.5)},
    decorate, decoration={snake, segment length=1mm, amplitude=0.2mm}},
  wedgelower/.style n args={6}{insert path={
      (0, .5) node[feynman/labelleft]{$a$} node[feynman/node #1]
      (0,-.5) node[feynman/labelleft]{$b$} node[feynman/node #2]
      (1.8, .5) node[feynman/labelright]{$1$} node[feynman/node #3]
      (1.8,-.5) node[feynman/labelright]{$2$} node[feynman/node #4]
      (0.9, .1) node[feynman/labelupper]{$#5$}
      (0.9,-.5) node[feynman/labellower]{$#6$}
      (0,-.5) -- (1.8, .5)
      (0,-.5) -- (1.8, -.5)},
    decorate, decoration={snake, segment length=1mm, amplitude=0.2mm}}
}
\newcommand*\tfWedgeUpper[1][]{\tfInt{wedgeupper}{#1}}
\newcommand*\tfWedgeLower[1][]{\tfInt{wedgelower}{#1}}

\usepackage{tikzit}
\usetikzlibrary{arrows.meta}
\input{tikzit/styles.tikzstyles}

\begin{document}

   \title{Wigner interferometry}


%

   \author{Vladimir Lenok}

   \institute{Fakult\"{a}t f\"{u}r Physik, Universit\"{a}t Bielefeld, Postfach 100131, 33501 Bielefeld, Germany\\
   \email{vlenok@physik.uni-bielefeld.de}}

   \date{Received September XX, 20XX}


  \abstract
  {}
   {This work aims to introduce a new method of interferometric measurements based on correlation of Wigner functions constructed for the fields from a distant source.}
   {This work presents theoretical studies of correlation of the Wigner functions.
    Numerical simulations support the findings.}
   {It is shown that in comparison to the correlation of the fields, a correlation of their Wigner functions samples twice higher spatial frequencies of the source intensity distribution and has about twice smaller scale of the spatial pattern.
     This opens a fundamental possibility to improve the angular resolution of interferometric measurements by a factor of two.}
  {}


   \maketitle

\section{Introduction}
Radio interferometry is the key technique in modern radio astronomy for imaging distant objects in space with high angular resolution, which provides invaluable information about their morphology as well as astrophysical processes occurring both within the objects themselves and in their proximity.
After a decades-long history of conceptual and technological development, this technique is now the standard imaging method in currently operating telescopes with synthetic apertures like the Karl G. Jansky Very Large Array~\citep{vla-main}, the MeerKAT radio telescope~\citep{meerkat-main}, and the International Low-Frequency Array~\citep{lofar-main}, and is also a fundamental operation principle for future large-scale systems such as the Square Kilometre Array~\citep{ska-main}.

To achieve resolutions exceeding capabilities of individual telescopes, multiple such telescopes can be combined to a single synthetic aperture system with a transcontinental scale by means of the very-long baseline interferometry (VLBI) technique~\citep{thompson2017}.
Traditionally, this technique is used to observe active galactic nuclei, their jets, and other compact objects poorly resolvable by connected-type interferometers enabling studies of astrophysical processes in such highly energetic regions~\citep{m87-review, agn-new-targets} and providing key measurements for geodesy and astrometry~\citep{itrf, icrf}.
Recent progress in this field enabled imaging of the black hole shadows~\citep{eht-m87-1, eht-cagA-1}, detecting fine morphology of the blazar OJ~287~\citep{radioastron-agn}, and resolving a water maser NGC~4258 at record angular resolution~\citep{radioastron-megamaser}.

From a conceptual point of view, a radio interferometer operates as pairs of spatially separated antennas, so-called baselines, measuring pairwise correlations of the electromagnetic field~\citep{thompson2017}.
These measurements form an observed sample of the coherence function.
Even though it is known that the coherence function is connected to the intensity distribution over the sky via a Fourier transform, reconstruction of this distribution constitutes a complex problem.
The complexity originates from the fact that the observed sample is limited in coverage and goes only up to a certain maximum baseline distance.
The latter defines a natural ``diffraction'' limit to the angular resolution of an interferometric system at a given wavelength.
There are several algorithms capable of solving this problem and reconstructing the image with an angular resolution limited by the diffraction limit or even better (see~\citet{eht-imaging} for a discussion about existing approaches to the source imaging and practical aspects of their application; see~\citet{alma-super-resolution} for a notable example of ``super resolution'' observations; see~\citet{lofar-subarc-imaging} and~\citet{mightee} for typical examples of imaging procedures).

However, regardless of any particular approach or algorithms, it is clear that from the physics perspective, the final resolution is driven by the amount of the observed spatial information from the sky and, in particular, by the greatest observed spatial frequency of the sky intensity distribution defined by the maximum baseline.
The problem of the greatest observed spatial frequency is not purely instrumental.
It originates from the correlation properties of the field.
Namely, from the fact that the sky image and the coherence pattern are connected via a Fourier transform~\citep{thompson2017, Born_Wolf}.

This paper proposes to measure a different quantity to address the classical limit of the angular resolution.
Instead of correlation of the fields themselves, we can construct Wigner functions from them and correlate these functions instead.
We can refer to this method as ``Wigner interferometry.''
Analytical computations reveal that under reasonable assumptions regarding the incoming radiation the correlation properties of the Wigner functions differ from those known for the fields themselves.
The main difference relevant for radio interferometry is the fact that at a given baseline the new spatial correlation pattern contains twice higher spatial frequencies of the source intensity distribution in comparison to the correlation of the fields themselves.
As a consequence, this leads to about a twice smaller coherence pattern in terms of scale.
Numerical simulations reproduce this effect.
This finding opens a fundamental possibility to double the angular resolution of interferometric imaging of distant objects.

This paper summarizes results of the initial theoretical investigations of correlations of Wigner functions in application to radio interferometry.
To simplify the computations, we consider only a single polarization of the fields, which, however, does not affect the general nature of the observed effects.
Also, to enable a fully analytical solution of the problem, we assume that the incoming radiation is a stationary zero-mean Gaussian process with a bandlimited white spectrum.
The assumption of Gaussianity appears to be reasonable since the most of the radio emission observed from the sources in space is the synchrotron emission~\citep{condon-essential-radio-astronomy}, which is known to be a Gaussian process~\citep{stat-synchrotron}.
The assumed spectrum can be straightforwardly achieved in actual observations by whitening of the incoming radiation similar to a way it is done for the gravitation wave detection~\citep{ligo-guide}.

The organization of this paper is as follows.
Section~\ref{section-established-methods} briefly recalls fundamentals of two established methods of interferometric measurements.
Section~\ref{section-two-points} considers a correlation of the Wigner functions for the case of a source consisting of only two radiating points in order to trace down non-vanishing contributions.
Section~\ref{section-generalization} generalizes these results to a continuous source.
It turns out that there are two specific types of the contribution to the resulting coherence pattern, which at least in principle can have physical interpretation.
Section~\ref{section-bandlimited-source} considers the case of a band-limited white spectrum to obtain equations for a source of a generic shape.
Finally, Section~\ref{section-exact-solutions} presents two exact solutions for square and circular sources and shows a comparison of them against numerical simulations, which are briefly described in Section~\ref{section-simulation}.

\section{Established methods}
\label{section-established-methods}
The vast majority of sources in space and cosmos produce spatially non-coherent radio emission.
However, when radiation from them arrives to the Earth, it is coherent, and this fact enables interferometric observations of distant objects in space.

This fundamental property of radiation to acquire spatial coherence while propagating away from a non-coherent source is summarized by a theorem named after Pieter Hendrik van Cittert and Frits Zernike.
If we consider a distant non-coherent source with an angular distribution of intensity $I(\chi, \psi)$, this theorem states that the value of coherence of radiation between the points $P_1$ and $P_2$ (see Fig.~\ref{fig:geometry}) is a Fourier transform \citep{Born_Wolf} 
\begin{equation}
  \label{eq-intro-vcz}
  \langle V_{1}(t) V_{2}(t) \rangle =
  \iint\limits_{\sigma} I(\chi, \psi) \mathrm{e}^{-ik(p\chi + q\psi)} \,\mathrm{d}\chi\,\mathrm{d}\psi,
\end{equation}
where $\chi$, $\psi$ are corresponding angular coordinates of the source ($\chi \approx x/R$, $\psi \approx y/R$), $p = X_1 - X_2$ and $q = Y_1 - Y_2$ are distances between the points $P_1$ and $P_2$ along x and y axes correspondingly (see Appendix~\ref{appendix-VV} for derivation of this equation).
Here, we also straightforwardly see that $p$ and $q$ are spatial frequencies of the source intensity distribution since they are conjugate variables to the angular coordinates $\chi$ and $\psi$ in the Fourier transform.
Equation~\ref{eq-intro-vcz} is a foundation for radio and optical/infrared interferometry (see \citet{thompson2017} and \citet{optical-interferometry} correspondingly).

Correlations of field fluctuations is not the only quantity which can be measured and used for observation of distant objects.
Instead of correlation of the fields themselves, we can consider a correlation of their intensities
\begin{equation}
  \langle I_1(t) I_2(t) \rangle =
  \langle V_1(t) V_1(t) V_2(t) V_2 (t) \rangle.
\end{equation}
If we carry out computations for a noncoherent source as above, we arrive at an equation
\begin{equation}
  \label{eq-intro-II}
  \langle I_1(t) I_2(t) \rangle = \frac{1}{4} \bar{I}_1 \bar{I}_2 + 2 \left[\langle V_1(t) V_2(t) \rangle\right]^2,
\end{equation}
where the quantity $\langle V_1(t) V_2(t) \rangle$ is the same as in Eq.~(\ref{eq-intro-vcz}).
This equation is a foundation for the intensity interferometry invented by Robert Hanbury Brown and developed together with Richard Q. Twiss (see \citet{hanbury-brown} for the historical account and bibliography of the original works).
A detailed derivation of Eqs.~(\ref{eq-intro-vcz}) and (\ref{eq-intro-II}) one can find in Appendices \ref{appendix-VV} and \ref{appendix-II} correspondingly.

We can see that even though correlation of the field fluctuations and correlation of the field intensities form two distinct types of interferometry, they measure essentially the same physical quantity.

\section{A case of two radiating points}
\label{section-two-points}

Let us consider a new object: a correlation of the Wigner functions, or the Wigner time-frequency distributions, constructed for the fields arrived from a distant source.
Formally, we define this object as
\begin{multline}
  \label{eq:w1w2}
  W(P_1, P_2, P_a, P_b) =
  \langle W_1(t) W_2(t)\rangle \\ =
  \iint\limits_{-\infty}^{+\infty}
  \langle
  V_1\left(t + \xi/2\right) V_1\left(t - \xi/2\right)
  V_2\left(t + \eta/2\right) V_2\left(t - \eta/2\right)
  \rangle \\
  \times
  ~\mathrm{e}^{-i\omega\xi}
  ~\mathrm{e}^{-i\omega\eta}
  ~\mathrm{d} \xi
  ~\mathrm{d} \eta,
\end{multline}
where $\omega$ denotes a given frequency and the averaging is the ensemble average.
To be concise, we will refer this object to as Wigner correlation function.

Our first goal is to understand what kind of contributions form the resulting correlation function of the Wigner distribution of the radiation.
The simplest case, which enables analytical investigation of this matter, is a distant source made of only two radiating points.
We consider it in this section.

\begin{figure}
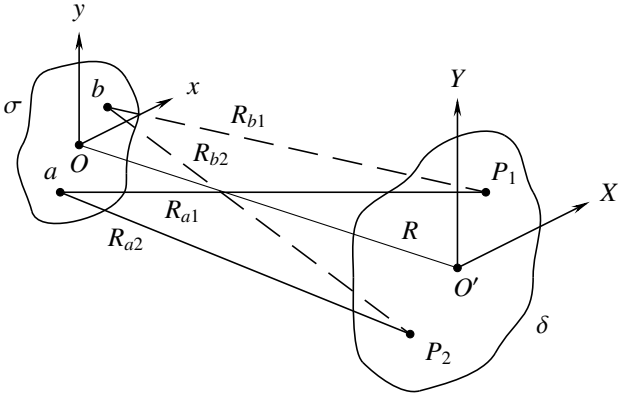

\ctikzfig{tikzit/coordinates}
\caption{An illustration of used geometrical quantities.}
\label{fig:geometry}
\end{figure}

Figure~\ref{fig:geometry} illustrates the case of two radiation points with the relevant geometrical quantities.
Field vibrations, if we use the terminology from~\citet{Born_Wolf}, from a point $a$ at a point $P_1$ of the detection plane has a form of spherical waves
\begin{equation}
\label{eq:singleV}
  V_{a1} = \frac{1}{R_{a1}}
  A_a \left(t - R_{a1}/c\right)\,
  \cos \left[
    \omega \left( t - R_{a1}/c \right)
    \right].
\end{equation}
The radiation from the point $a$ to the point $P_2$, and from the point $b$ to the points $P_1$ and $P_2$ have identical forms with a corresponding change of the distances and amplitudes.

For the case of two points, it is straightforward to write the radiation at the points $P_1$ and $P_2$ explicitly.
It is simply sums of the radiation from both points
\begin{align}
  \label{eq:Vsum}
V_1 &= V_{a1} + V_{b1}, \\
V_2 &= V_{a2} + V_{b2}.
\end{align}

The relevant information about the radiation from the source is contained in the fourth-order correlation function in Eq.~(\ref{eq:w1w2}).
For convenience, we denote this function as
\begin{equation}
  \label{eq-gamma}
  \gamma =
  \langle
  V_1\left(t + \xi/2\right) V_1\left(t - \xi/2\right)
  V_2\left(t + \eta/2\right) V_2\left(t - \eta/2\right)
  \rangle.
\end{equation}
In the case of the two-point source, this function has a form
\begin{equation}
  \begin{aligned}
  \label{eq:VVVVsums}
  \gamma
  =
  \langle
  &\left[
  V_{a1}\left(t + \xi/2\right) + V_{b1}\left(t + \xi/2\right)
  \right] \\
  &\left[
  V_{a1}\left(t - \xi/2\right) + V_{b1}\left(t - \xi/2\right)
  \right] \\
  &\left[
  V_{a2}\left(t + \eta/2\right) + V_{b2}\left(t + \eta/2\right)
  \right] \\
  &\left[
  V_{a2}\left(t - \eta/2\right) + V_{b2}\left(t - \eta/2\right)
  \right]
  \rangle.
  \end{aligned}
\end{equation}
where the radiation at each point is the sum of two contributions as in Eq.~(\ref{eq:Vsum}) with corresponding arguments.
It is convenient to introduce a shorthand notation for these terms.
Let $a_1(+\xi)$ denote $V_{a1}\left(t + \xi/2\right)$.
The letter is the point of the radiation origin and
the subscript is the point of observation.
The letter and its sign in the brackets denote the corresponding time lag and its sign in the argument.
In this notation, the fourth order correlation function (\ref{eq:VVVVsums}) simplifies to
\begin{multline}
  \gamma =
  \langle
  \left[a_1(+\xi)  + b_1(+\xi) \right]
  \left[a_1(-\xi)  + b_1(-\xi) \right] \\
  \left[a_2(+\eta) + b_2(+\eta)\right]
  \left[a_2(-\eta) + b_2(-\eta)\right]
  \rangle.
\end{multline}
If we expand the brackets now, we end up at a sum of twelve terms.
The first four of them are
\begin{equation}
\begin{aligned}
  \label{eq:VVVVsums-simplified}
  \gamma =
  &\langle
  a_1(+\xi)\, a_1(-\xi)\, a_2(+\eta)\, a_2(-\eta)
  \rangle \\ 
  +
  &\langle
  a_1(+\xi)\, a_1(-\xi)\, a_2(+\eta)\, b_2(-\eta)
  \rangle \\
  +
  &\langle
  a_1(+\xi)\, a_1(-\xi)\, b_2(+\eta)\, a_2(-\eta)
  \rangle \\
  +
  &\langle
  a_1(+\xi)\, a_1(-\xi)\, b_2(+\eta)\, b_2(-\eta)
  \rangle + \dots
\end{aligned}
\end{equation}
The remaining three groups also contain four terms each with the same terms on the positions three and four, but starting with
$\langle a_1(+\xi)\, b_1(-\xi) \dots \rangle$, $\langle b_1(+\xi)\, a_1(-\xi) \dots \rangle$, and $\langle b_1(+\xi)\, b_1(-\xi) \dots \rangle$
correspondingly.

To be able to expand the fourth-order correlation functions appearing in~(\ref{eq:VVVVsums-simplified}), we should make some assumptions regarding the radiation we observe.
In this work, we assume that the radiation is a zero-mean stationary Gaussian ergodic process.
This assumption appears to be reasonable since most of the emission observed in radio astronomy is synchrotron emission known to be a Gaussian random process~\citep{condon-essential-radio-astronomy, stat-synchrotron}.
This assumption allows us to use the Isserlis's probability theorem, which has a following form for our case~\citep{isserlis-theorem}
\begin{equation}
  \langle x_1 x_2 x_3 x_4 \rangle =
  \langle x_1 x_2 \rangle \langle x_3 x_4 \rangle +
  \langle x_1 x_3 \rangle \langle x_2 x_4 \rangle +
  \langle x_1 x_4 \rangle \langle x_2 x_3 \rangle.
\end{equation}
When we apply this theorem to the four-order correlation function~(\ref{eq:VVVVsums-simplified}), each of the twelve terms forms three new terms.
Thus, there are 36 terms in total.

Despite such a large number of terms, we will see in a moment that only few of them are relevant.
First, the two points $a$ and $b$ are noncoherent.
This condition leads to vanishing all the terms that are not of the form $\langle aaaa \rangle$, $\langle bbbb \rangle$, or $\langle aabb \rangle$ (with any positions of a pair of $a$ and a pair of $b$).
For example, the second term in (\ref{eq:VVVVsums-simplified}) after application of the Wick's theorem becomes
\begin{multline}
  \langle
  a_1\, a_1\, a_2\, b_2
  \rangle =
  \langle a_1\, a_1 \rangle \langle a_2\, b_2 \rangle +
  \langle a_1\, a_2 \rangle \langle a_1\, b_2 \rangle \\ +
  \langle a_1\, b_2 \rangle \langle a_1\, a_2 \rangle = 0,
\end{multline}
where we omitted the arguments for clarity.
All correlations of the form $\langle a\,b \rangle$ are zeros here due to non-coherence of the points.
The same happens with the other terms.

We can still eliminate some of the remaining term.
To see it, let us consider a following term from Eq.~(\ref{eq:VVVVsums-simplified}) expanded with the Isserlis's theorem
\begin{multline}
    \langle a_1\, b_1\, b_2\, a_2 \rangle
  =  \langle a_1\, b_1  \rangle
     \langle b_2\, a_2  \rangle
  + \langle a_1\, b_2 \rangle
     \langle b_1\, a_2  \rangle \\
  +  \langle a_1\, a_2 \rangle
     \langle b_1\, b_2  \rangle
  =  \langle a_1\, a_2 \rangle
     \langle b_1\, b_2  \rangle
\end{multline}
Here, we omitted again the arguments for clarity.
If we analyze the appeared terms, it is clear that the first two of them are zeros due to the noncoherence of the source points, but the last term survives under this condition.
Let us see what happens with this last term in the integral in Eq.~(\ref{eq:w1w2}) defining the Wigner coherence.
If we insert this term in the integral, we obtain
\begin{equation}
  \label{eq:Jinit}
  \iint\limits_{-\infty}^{+\infty}
  \langle a_1(+\xi)\,  a_2(-\eta) \rangle
  \langle b_1(-\xi)\,  b_2(+\eta) \rangle
  ~\mathrm{e}^{-i\omega\xi}
  ~\mathrm{e}^{-i\omega\eta}~
  ~\mathrm{d} \xi
  ~\mathrm{d} \eta = J.
\end{equation}
The source emits stationary radiation.
Thus, the correlation functions are functions of a sum $\xi + \eta$.
Due to this fact, the integral can be represented as
\begin{equation}
  J =
  \int\limits_{-\infty}^{+\infty}
  \int\limits_{-\infty}^{+\infty}
  f(\eta + \xi)
  ~\mathrm{e}^{-i\omega\xi}
  ~\mathrm{e}^{-i\omega\eta}~
  \,\mathrm{d} \xi
  \,\mathrm{d} \eta.
\end{equation}
If we change the variables to $\xi' = \xi + \eta$ and $\eta' = \eta - \xi$, which corresponds to a $\pi/4$-turn of the coordinate system, we arrive at a following form that integral
\begin{equation}
  J =
  2
  \int\limits_{-\infty}^{+\infty}
  ~\mathrm{d} \eta'
  \int\limits_{-\infty}^{+\infty}
  ~\mathrm{d} \xi'
  ~\mathrm{e}^{-i\omega\eta'}~
  f(\xi'),
\end{equation}
which clearly diverges.
The origin and nature of this divergence is clear: the integrand depends only on $\xi' = \xi + \eta$.

There could be multiple ways to regularize such a divergence.
It seems that the most natural way to do it is to use the geometry of the integrand.
Let us first, represent the integral as a limit of a following integral over a band going along $\eta = \xi$ line
\begin{equation}
  J =
  \lim_{\eta_0 \to \infty}
  \int\limits_{-\infty}^{\infty}
  \int\limits_{\xi - \eta_0}^{\xi + \eta_0}
  f(\xi + \eta)
  \,\mathrm{d}\eta
  \,\mathrm{d}\xi.
\end{equation}
And now we introduce a normalization over a width of the band in a similar fashion as in construction of average values
\begin{equation}
  J' =
  \lim_{\eta_0 \to \infty}
  \frac{1}{4\eta_0}
  \int\limits_{-\infty}^{\infty}
  \int\limits_{\xi - \eta_0}^{\xi + \eta_0}
  f(\xi + \eta)
  \,\mathrm{d}\eta
  \,\mathrm{d}\xi
\end{equation}
If we make the same change of variables as above, we arrive at
\begin{equation}
\label{eq:Jnormalization}
  J' =
  \lim_{\eta_0 \to \infty} \frac{1}{2\eta_0}
  \int\limits_{-\eta_0}^{\eta_0} \, \mathrm{d}\eta'
  \int\limits_{-\infty}^{\infty} \, \mathrm{d}\xi'
  f(\xi') =
    \int\limits_{-\infty}^{\infty}
  f(\xi')
  \,\mathrm{d}\xi'
\end{equation}
constituting a non-diverging version of the integral (\ref{eq:Jinit}).

If we analyze all the pairwise contributions in Eq.~(\ref{eq:VVVVsums-simplified}), we notice that only a few of them are diverging.
These are exactly the terms that do not vanish after the regularization described above and constitute the contributions relevant for computing the Wigner correlation.
In the considered case of the two-point source, there are four such terms
\begin{equation}
\begin{aligned}
  \label{eq:J2point}
  \gamma &= \langle a_1(+\xi) a_2(-\eta) \rangle \langle a_1(-\xi) a_2(+\eta) \rangle \\
    &+ \langle a_1(+\xi) a_2(-\eta) \rangle \langle b_1(-\xi) b_2(+\eta) \rangle \\
    &+ \langle b_1(+\xi) b_2(-\eta) \rangle \langle a_1(-\xi) a_2(+\eta) \rangle \\
    &+ \langle b_1(+\xi) b_2(-\eta) \rangle \langle b_1(-\xi) b_2(+\eta) \rangle.
\end{aligned}
\end{equation}

Now when it is clear what kind contributions are relevant, we can generalize this result to a case of a continuous extended source.

\section{Generalization to a continues source}
\label{section-generalization}
Let us represent the correlation function (\ref{eq:J2point}), which we just obtained, in a following diagrammatic form
\begin{equation}
\begin{aligned}
  \gamma &=\biggl< \tfWedgeUpper{****}{+\xi}{-\eta} \biggl>\,
         \biggl< \tfWedgeUpper{****}{-\xi}{+\eta} \biggl>
       \,+\,\biggl< \tfWedgeUpper{****}{+\xi}{-\eta} \biggl>\,
         \biggl< \tfWedgeLower{****}{-\xi}{+\eta} \biggl> \\
       &+\biggl< \tfWedgeLower{****}{+\xi}{-\eta} \biggl>\,
         \biggl< \tfWedgeUpper{****}{-\xi}{+\eta} \biggl>
       \,+\,\biggl< \tfWedgeLower{****}{+\xi}{-\eta} \biggl>\,
         \biggl< \tfWedgeLower{****}{-\xi}{+\eta} \biggl>.
\end{aligned}
\end{equation}
If we observe carefully, we see that there are two types of contributions: those which are products of the correlation function originating from the two different points (the second and third terms), and self-product contributions (the first and fourth terms).
From this observation, we can heuristically see how to obtain an extension of the obtained result to a case of a continuous source of an arbitrary shape.
We can denote as $\gamma_{ab}$ a contribution of the first type
\begin{equation}
  \gamma_{ab} =
  \biggl< \tfWedgeUpper{****}{+\xi}{-\eta} \biggl>\,
  \biggl< \tfWedgeLower{****}{-\xi}{+\eta} \biggl>
\end{equation}
and as $\gamma_{aa}$ of the second type
\begin{equation}
  \gamma_{aa} =
  \biggl< \tfWedgeUpper{****}{+\xi}{-\eta} \biggl>\,
  \biggl< \tfWedgeUpper{****}{-\xi}{+\eta} \biggl>,
\end{equation}
where now the locations of the points $a$ and $b$ are arbitrary on the source $\sigma$ (see Fig.~\ref{fig:geometry}).
To obtain correlations from a continues source, we should integrate them over the source surface.
The $\gamma_{ab}$
should be integrated twice (for the point $a$ and for the point $b$) and the contribution $\gamma_{aa}$ only once.
Namely,
\begin{multline}
  \label{eq:w1w2cont}
  \langle W_1(t) W_2(t) \rangle = W_{ab} + W_{aa} \\
  = \int\limits_{\sigma}
  \int\limits_{\sigma}
  J'_{ab}(\sigma_a, \sigma_b)
  ~\mathrm{d}\sigma_a ~\mathrm{d} \sigma_b
  +
  \int\limits_{\sigma}
  J'_{aa}(\sigma_a)
  ~\mathrm{d}\sigma_a,
\end{multline}
where
\begin{align}
\label{eq:Jab-general}
  J'_{ab}(\sigma_a, \sigma_b) &=
  \lim_{\eta_0 \to \infty}
  \frac{1}{4\eta_0}
  \int\limits_{-\infty}^{\infty}
  \int\limits_{\xi - \eta_0}^{\xi + \eta_0}
  \gamma_{ab}(\xi + \eta)
  \,\mathrm{d}\eta
  \,\mathrm{d}\xi, \\
  \label{eq:Jaa-general}
  J'_{aa}(\sigma_a) &=
  \lim_{\eta_0 \to \infty}
  \frac{1}{4\eta_0}
  \int\limits_{-\infty}^{\infty}
  \int\limits_{\xi - \eta_0}^{\xi + \eta_0}
  \gamma_{aa}(\xi + \eta)
  \,\mathrm{d}\eta
  \,\mathrm{d}\xi
\end{align}
are the regularized integrals, which give finite results, and $\mathrm{d}\sigma_a$, $\mathrm{d} \sigma_b$ are source elements corresponding locations of the points $a$ and $b$.

It appears to seem difficult to obtain analytic solutions directly for Eqs.~(\ref{eq:Jab-general}) and (\ref{eq:Jaa-general}).
However, we can obtain simplified forms of them by considering a general structure of $\gamma_{ab}$ and $\gamma_{aa}$ products for which we can find analytic solutions.

Let us consider a product between correlation functions originating from two arbitrary points $a$ and $b$ on the source
\begin{equation}
  \label{eq:gamma-ab-general}
  \gamma_{ab} =
  \langle a_1(+\xi) a_2(-\eta) \rangle \langle b_1(-\xi) b_2(+\eta) \rangle.
\end{equation}
The first correlation function has the following general form (see Fig.~\ref{fig:geometry} for geometrical quantities)
\begin{multline}
\label{eq:aa-general-form}
  \langle a_1(+\xi) a_2(-\eta) \rangle =
  \frac{1}{R_{a1}}
  \frac{1}{R_{a2}}
  \biggl<
  A_a\left(t + \frac{\xi}{2} - \frac{R_{a1}}{c} \right)
  A_a\left(t - \frac{\eta}{2} - \frac{R_{a2}}{c} \right)\biggr> \\
  \times
  \mathrm{Re}\left\{
  \mathrm{e}^{-i \omega \left(
    t + \xi/2 - R_{a1}/c
    \right)}
  \mathrm{e}^{i \omega \left(
    t - \eta/2 - R_{a2}/c
    \right)}
  \right\}.
\end{multline}
Here, to average the product of the cosine functions originating from the original form for the radiation (\ref{eq:singleV}), we express them as corresponding complex exponentials, let us denote them as $z_1$ and $z_2$, and take the real part of a product $\mathrm{Re}\{ z_1 z_2^*\}$~\citep[see][chap.~6, \S 48]{landau2}.

The remaining correlation function for a general case is
\begin{multline}
  \biggl<
  A_a\left(t + \frac{\xi}{2} - \frac{R_{a1}}{c} \right)
  A_a\left(t - \frac{\eta}{2} - \frac{R_{a2}}{c} \right)\biggr> \\ =
  I(\sigma_a) R\left( -\frac{\xi + \eta}{2} + \frac{1}{c}\left( R_{a1} - R_{a2} \right) \right)
\end{multline}
where $I(\sigma_a)$ denotes the intensity per unit area of the source at point $a$, and $R$ denotes the correlation function of the radiation originating from the point $a$.
If we use the general form of the correlation function (\ref{eq:aa-general-form}) and write the real part explicitly, we obtain
\begin{multline}
\label{eq:aa-subgeneral-form}
  \langle a_1(+\xi) a_2(-\eta) \rangle =
  \frac{I(\sigma_a)}{R_{a1}R_{a2}}
  R\left( -\frac{\xi + \eta}{2} + \frac{R_{a1} - R_{a2}}{c} \right) \\
  \times
  \cos\left[ \omega \frac{\xi + \eta}{2} - \omega \frac{R_{a1} - R_{a2}}{c} \right].
\end{multline}

By following the very same logic, we arrive at the following equation for the second correlation function in Eq.~(\ref{eq:gamma-ab-general})
\begin{multline}
\label{eq:bb-subgeneral-form}
  \langle b_1(-\xi) b_2(+\eta) \rangle =
  \frac{I(\sigma_b)}{R_{b1}R_{b2}}
  R\left( \frac{\xi + \eta}{2} + \frac{R_{b1} - R_{b2}}{c} \right) \\
  \times
  \cos\left[ \omega \frac{\xi + \eta}{2} + \omega \frac{R_{b1} - R_{b2}}{c} \right].
\end{multline}

Now we can clearly see that both correlation functions depend on $\xi + \eta$.

If we use the obtained forms of the correlations (\ref{eq:aa-subgeneral-form}) and (\ref{eq:bb-subgeneral-form}) in Eq.~(\ref{eq:gamma-ab-general}) and consider its integral (\ref{eq:Jab-general}), we obtain a general form of this integral by changing variables $\xi' = \xi + \eta$ and $\eta' = \eta - \xi$ as we did above and use the regularization (\ref{eq:Jnormalization}) for simplification
\begin{multline}
\label{eq:cross-inter-general}
  J'_{ab}(\sigma_a, \sigma_b) =
  \frac{I(\sigma_a)}{R_{a1}R_{a2}}
  \frac{I(\sigma_b)}{R_{b1}R_{b2}}
  \int\limits_{-\infty}^{\infty}
  R\left( -\frac{\xi'}{2} + \alpha \right)
  R\left( \frac{\xi'}{2} + \beta \right)\\
  \times
  \cos\left[ \frac{\omega \xi'}{2} - \omega \alpha \right]
  \cos\left[ \frac{\omega \xi'}{2} + \omega \beta \right]
  \,\mathrm{d}\xi'.
\end{multline}
The constants $\alpha$ and $\beta$ denote following time intervals
\begin{align}
  \alpha &= \frac{1}{c} (R_{a1} - R_{a2}),\\
  \beta  &= \frac{1}{c} (R_{b1} - R_{b2}).
\end{align}
The distances here are functions of locations of the points $P_1$, $P_2$, $a$, and $b$.
Therefore, the functions $\alpha$ and $\beta$ are also functions of these locations.

A simpler form of $J'_{aa}(\sigma_a)$ can be written right away from Eq.~(\ref{eq:cross-inter-general}) by substituting all the quantities for the point $b$ with those for the point $a$
\begin{multline}
\label{eq:self-inter-general}
  J'_{aa}(\sigma_a) =
  \frac{I^2(\sigma_a)}{R_{a1}^2R_{a2}^2}
  \int\limits_{-\infty}^{\infty}
  R\left( -\frac{\xi'}{2} + \alpha \right)
  R\left( \frac{\xi'}{2} + \alpha \right)\\
  \times
  \cos\left[ \frac{\omega \xi'}{2} - \omega \alpha \right]
  \cos\left[ \frac{\omega \xi'}{2} + \omega \alpha \right]
  \,\mathrm{d}\xi'.
\end{multline}

The obtained Eqs.~(\ref{eq:cross-inter-general}) and (\ref{eq:self-inter-general}) are simplified general forms of Eqs.~(\ref{eq:Jab-general}) and (\ref{eq:Jaa-general}).

Thus, to compute the Wigner correlation in a general continuous case, we should
\begin{enumerate}
\item select a specific correlation function $R(\tau)$,
\item find the cross-product and self-product contributions by using Eqs.~(\ref{eq:cross-inter-general}) and (\ref{eq:self-inter-general}),
\item integrate both contributions over the source surface and sum up the results as in Eq.~(\ref{eq:w1w2cont}).
\end{enumerate}

The correlation function (\ref{eq:w1w2}) is not normalized.
It does not impose any difficulties on the theory, however, for practical situations, it seems reasonable to consider a normalized form of it.
For example, a form normalized to the peak value of the correlation
\begin{equation}
    w = \frac{\langle W_1(t, \omega) W_2(t, \omega)\rangle}
             {\langle W_1(t, \omega) W_1(t, \omega)\rangle},
\end{equation}
which by construction resembles the complex degree of coherence used in coherence theory (see for example~\citet{Born_Wolf}).

It seems that the general form of the Wigner correlation function does not provide better insights regarding its properties.
To reveal some of the properties, we should assume a particular model of the correlation function $R(\tau)$.

\section{A source with a band-limited white spectrum}
\label{section-bandlimited-source}

As an example we consider a source emitting a radiation with a band-limited white spectrum centered at the frequencies $\pm\omega_0$ and the bandwidth of $2B$.
The corresponding power spectrum is
\begin{equation}
  \label{eq:power-spectrum}
  G(\omega) = \begin{cases}
    1 & \text{ when $\omega_0 - B \leq |\omega| \leq \omega_0 + B$ },\\
    0 & \text{ when $\omega_0 - B > |\omega| > \omega_0 + B$}.
    \end{cases}
\end{equation}
Equivalently, this situation corresponds to a receiver, which detects only the denoted frequency band from a white noise source.

On the one hand, this case is simple enough to enable fully analytical solutions.
On the other hand, it is close to various real situations.

\subsection{Cross-product contribution}
The correlation function for such a Gaussian white noise process with the band-limited spectrum (\ref{eq:power-spectrum}) is known to be~\citep[see][chap.~5, table~5.1]{bendat}
\begin{equation}
  \label{eq:R}
  R(\tau) = \frac{\sin B\tau}{B\tau} \cos \omega_0 \tau.
\end{equation}
Let us find the cross-product contribution $W_{ab}$ with this correlation function.
For clarity, we use explicit Eqs.~(\ref{eq:aa-subgeneral-form}) and (\ref{eq:bb-subgeneral-form}) for individual correlations instead of the final Eq.~(\ref{eq:cross-inter-general}).

The resulting equation for correlaiton (\ref{eq:aa-subgeneral-form}) is
\begin{multline}
  \label{eq:a1a2}
  \langle a_1(+\xi) a_2(-\eta) \rangle =
  \frac{I(\sigma_a)}{R_{a1}R_{a2}}
  \frac{\sin\left[ B \left(\xi + \eta\right)/2 - B\alpha \right]}{B\left(\xi + \eta\right)/2 - B\alpha} \\
  \times
  \cos\left[ \omega_0 \frac{\xi + \eta}{2} - \omega_0 \alpha \right]
  \cos\left[ \omega \frac{\xi + \eta}{2} - \omega \alpha \right].
\end{multline}
The resulting equation for the correlation (\ref{eq:bb-subgeneral-form}) is nearly identical
\begin{multline}
  \label{eq:b1b2}
  \langle b_1(-\xi) b_2(+\eta) \rangle =
  \frac{I(\sigma_b)}{R_{b1}R_{b2}}
  \frac{\sin\left[ B \left(\xi + \eta\right)/2 + B\beta \right]}{B\left(\xi + \eta\right)/2 + B\beta} \\
  \times
  \cos\left[ \omega_0 \frac{\xi + \eta}{2} + \omega_0 \beta \right]
  \cos\left[ \omega \frac{\xi + \eta}{2} + \omega \beta \right].
\end{multline}

\begin{figure}[t]
  \includegraphics{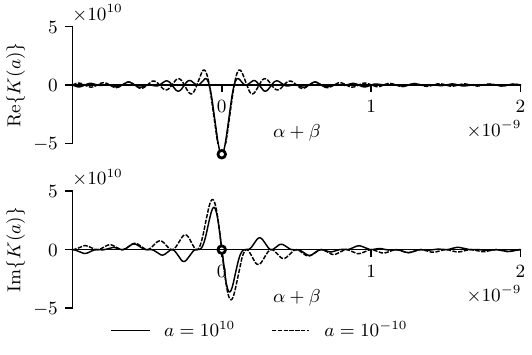}
  \caption{Real and imaginary values of the integral $K(a)$ about the point $\alpha + \beta = 0$ for $\omega = 2\pi \times 3 \times 10^9\,\mathrm{rad/s}$ and two values of the parameter $a$.
  The open dots are excluded due to removable discontinuities.}
  \label{fig:K}
\end{figure}

Now, we change variables $\xi' = \xi + \eta$ in both of these equations and insert them in Eq.~(\ref{eq:cross-inter-general}) to obtain a particular form of $J'_{ab}(\sigma_a, \sigma_b)$.
Then, the combination of sines and cosines in the integrand can be transformed by using standard trigonometric identities to obtain a linear combination of simpler integrals.
Thus after trivial but lengthy trigonometric transforms, we arrive at the following simplified form of the integral (\ref{eq:cross-inter-general})
\begin{multline}
  \label{eq:Jsum}
    J'_{ab} = \frac{4}{B^2}\frac{I(\sigma_a) I(\sigma_b)}{R_{a1}R_{a2}R_{b1}R_{b2}}
    \biggl[ A - B_1 + B_2 + B_3 - \frac{1}{2}(C_1 + C_2 \\ + C_3 + C_4 - C_5 - C_6)
    - \frac{1}{4}\left(D_1 + D_2 + D_3 + D_4\right) \biggr],
\end{multline}
where the individual terms are
\begin{align}
  A\phantom{_1}  &= \cos ((\alpha+\beta)B)\, \cos((\alpha+\beta)\omega)\, \cos((\alpha+\beta)\omega_0)\, K(0), \\
  B_1 &= \cos((\alpha+\beta)\omega)\, \cos((\alpha+\beta)\omega_0)\, K(B), \\
  B_2 &= \cos((\alpha+\beta)B)\, \cos((\alpha+\beta)\omega_0)\, K(\omega), \\
  B_3 &= \cos((\alpha+\beta)B)\, \cos((\alpha+\beta)\omega)\,  K(\omega_0), \\
  C_1 &= \cos((\alpha+\beta)\omega_0)\, K(\omega - B), \\
  C_2 &= \cos((\alpha+\beta)\omega_0)\, K(\omega + B), \\
  C_3 &= \cos((\alpha+\beta)\omega)\, K(\omega_0 - B), \\
  C_4 &= \cos((\alpha+\beta)\omega)\, K(\omega_0 + B), \\
  C_5 &= \cos((\alpha+\beta)B)\, K(\omega_0 - \omega), \\
  C_6 &= \cos((\alpha+\beta)B)\, K(\omega_0 + \omega), \\
  D_1 &= K(\omega_0 - \omega - B), \\
  D_2 &= K(\omega_0 - \omega + B), \\
  D_3 &= K(\omega_0 + \omega - B), \\
  D_4 &= K(\omega_0 + \omega + B),
\end{align}
and $K(a)$ is an identical integral depending on a parameter $a$\footnote{Since this parameter and the subscript $_a$ related to the point $a$ (see Fig.~\ref{fig:geometry}) are never in the same equations, this notation should not lead to a confusion.}
\begin{equation}
  \label{eq:Kint}
  K(a) = \int\limits_{-\infty}^{+\infty}
  \frac{\cos (a \xi' - a(\alpha - \beta))}{(\xi' - 2\alpha)(\xi' + 2\beta)}
  \mathrm{e}^{-i\omega \xi'}\,\mathrm{d}\xi'.
\end{equation}
Let us analyze this integral. 
Its solution in a sense of the Cauchy principal value can be straightforwardly obtained by using standard methods of complex integration 
\begin{multline}
  \label{eq:Kgeneral}
  K(a) = \frac{-i\pi}{4(\alpha+\beta)}
  \biggl[
    \mathrm{sgn}(\omega - a)\,
    \biggl(\mathrm{e}^{-i\left(\omega - a\right) 2 \alpha} -
           \mathrm{e}^{ i\left(\omega - a\right) 2 \beta } \biggr)
           \,\mathrm{e}^{-ia\left(\alpha-\beta\right)} \\
           +
    \mathrm{sgn}(\omega + a)\,
    \biggl(\mathrm{e}^{-i\left(\omega + a\right) 2 \alpha} -
    \mathrm{e}^{ i\left(\omega + a\right) 2 \beta } \biggr)
    \,\mathrm{e}^{ ia\left(\alpha-\beta\right)}
           \biggr].
\end{multline}
Figure~\ref{fig:K} shows real and imaginary parts of this solution for two very different values of the parameter $a$.
We can see that the real part has a maximum value about $\beta = -\alpha$ and falls very fast further away from it, and that the imaginary part has symmetric oscillations.
This behavior holds for a very large range of the values $a$.

Such a functional behavior indicates that a leading contribution to $J'_{ab}$ comes from ``interference'' of correlation functions of the source originating coming to the observer from the symmetric directions, or in other words, originating from symmetric areas of the source.

Let us use only this leading contribution to simplify $K(a)$.
Namely, the value of function at $\beta = -\alpha$.
The integral (\ref{eq:Kint}) for this case has a different form.
The poles merge into a second-order pole leading to a principle value, which is different from the one obtained in the general case (\ref{eq:Kgeneral})
\begin{equation}
\label{eq:K-for-Jab}
  K'(a) = -\frac{\pi}{2} \mathrm{e}^{-i2\omega\alpha}(|\omega - a| + |\omega + a|).
\end{equation}
By using this solution and a $\delta$-function, we can express the simplification formally as
\begin{equation}
  \label{eq:K-appox-delta}
  K(a) \approx K'(a)\,\delta(\alpha+\beta).
\end{equation}

Now we can use this approximation for integration over the source.
After assuming $\beta = -\alpha$, all the cosines in Eq.~(\ref{eq:Jsum}) can be greatly simplified.
They become unities and integration over the source should be done only for the function $K(a)$ itself.
Thus, the original Eq.~(\ref{eq:Jsum}) reduces to
\begin{equation}
\label{eq:Jab-simplified}
  J'_{ab} = \frac{4}{B^2}\frac{I(\sigma_a) I(\sigma_b)}{R_{a1}R_{a2}R_{b1}R_{b2}}
  \left[ K'_1 - \frac{1}{2}K'_2 - \frac{1}{4}K'_3\right],
\end{equation}
where
\begin{align}
  K'_1 &= K'(0) - K'(B) + K'(\omega) + K'(\omega_0), \\
  K'_2 &= K'(\omega - B) + K'(\omega + B) + K'(\omega_0 - B) \notag \\
      &+ K'(\omega_0 + B) - K'(\omega_0 - \omega) - K'(\omega_0 + \omega), \\
  K'_3 &= K'(\omega_0 - \omega - B) + K'(\omega_0 - \omega + B) \notag \\
      &+ K'(\omega_0 + \omega - B) + K'(\omega_0 + \omega + B).
\end{align}
Equation~(\ref{eq:Jab-simplified}) can be used to compute the cross-product contribution to the Wigner correlation function (\ref{eq:w1w2cont}).
Since one of the integrals disappears due to the $\delta$-function in the approximation~(\ref{eq:K-appox-delta}), the cross-product contribution is given by a following integral
\begin{equation}
  W_{ab} =
  \int\limits_{\sigma}
  J'_{ab}(\sigma_a)
  ~\mathrm{d}\sigma_a.
\end{equation}

If we look at the structure of Eq.~(\ref{eq:Jab-simplified}), we notice the integration affects only to the functions $K(a)$.
Thus, the problem reduces to a solution of the integral
\begin{equation}
\label{eq:K-integral}
  \int\limits_{\sigma}
  K'(a)
  ~\mathrm{d}\sigma_a.
\end{equation}

The function $\alpha$ appearing in the integral $K(a)$ (\ref{eq:K-for-Jab}) essentially holds entire information regarding the geometry of the problem.
If we follow the very same steps to simplify it for a very distant source as described in Appendix~\ref{appendix-VV} in derivation of the van Cittert-Zernike theorem, we arrive at
\begin{equation}
  \int\limits_{\sigma} K'(a) \,\mathrm{d}\chi\,\mathrm{d}\psi
  = f(a)\! \int\limits_{\sigma} I(\chi, \psi) I(-\chi, -\psi) \mathrm{e}^{i2 k \left( p\chi + q\psi \right)} \,\mathrm{d}\chi\,\mathrm{d}\psi,
\end{equation}
where $f(a)$ denotes a function
\begin{equation}
  f(a) = -\frac{\pi}{2}\, \mathrm{e}^{-i2\omega\zeta}\,(|\omega - a| + |\omega + a|).
\end{equation}
The remaining integral depends only on $\omega$ and the intensity distribution over the source.
Thus, it is identical for all the terms in Eq.~(\ref{eq:Jab-simplified}).
We denote this remaining integral as
\begin{equation}
\label{eq:L-general}
    L(\omega) =
    \int\limits_{\sigma} I(\chi, \psi) I(-\chi, -\psi) \mathrm{e}^{i2 k \left( p\chi + q\psi \right)} \,\mathrm{d}\chi\,\mathrm{d}\psi.
\end{equation}
Also, for convenience, we can denote all the remaining terms as a new function
\begin{equation}
  F(\omega, \omega_0, B) = f_1 - \frac{1}{2}f_2 - \frac{1}{4}f_3,
\end{equation}
where
\begin{align}
  f_1 &= f(0) - f(B) + f(\omega) + f(\omega_0), \\
  f_2 &= f(\omega - B) + f(\omega + B) + f(\omega_0 - B) \notag \\
      &+ f(\omega_0 + B) - f(\omega_0 - \omega) - f(\omega_0 + \omega), \\
  f_3 &= f(\omega_0 - \omega - B) + f(\omega_0 - \omega + B) \notag \\
      &+ f(\omega_0 + \omega - B) + f(\omega_0 + \omega + B).
\end{align}
By using the introduced notation, we can express the cross-product contribution to the Wigner correlation in a factorized form
\begin{equation}
\label{eq:Wab-final}
  W_{ab} =
  \frac{4}{B^2} \frac{1}{R^4} F(\omega, \omega_0, B)\, L(\omega).
\end{equation}

The information about geometry of the source and consequently about the resulting coherence pattern is contained in the function $L(\omega)$.
To find a particular analytical form of the cross-product contribution, we should solve the corresponding integral (\ref{eq:L-general}).
We will consider below two cases for which we can obtain fully analytical solutions.

Already at this point we can notice one important detail.
If we look at the integral for $L(\omega)$, we can notice that in comparison to the usual integral (\ref{eq-intro-vcz}) appearing the van Cittert-Zernike theorem the integral (\ref{eq:L-general}) for $L(\omega)$ contains an additional factor of two in the exponent.
This indicates that regardless of any particular distribution of intensity, the overall scale of the resulting coherence pattern is expected to be twice smaller than in the classical case.
We will see below that at least for the cases considered here, it is actually true.

\subsection{Self-product contribution}

Now, let us compute the contribution $W_{aa}$
in (\ref{eq:w1w2cont}).
To do it, we can repurpose the results for the cross-product case above.
Namely, we use Eq.~(\ref{eq:Jaa-general}) with
\begin{equation}
  \label{eq:gamma-aa-general}
  \gamma_{ab} =
  \langle a_1(+\xi) a_2(-\eta) \rangle \langle a_1(-\xi) a_2(+\eta) \rangle.
\end{equation}
The difference here from the previous case is that the point $b$ becomes the point $a$, which means $\beta = \alpha$ correspondingly.
Thus, we can immediately write a solution
\begin{multline}
  \label{eq:J2}
  J'_{aa}
  =
  \frac{4}{B^2}\frac{I^2(\sigma_a)}{R_{a1}^2 R_{a2}^2}
    \biggl[ A - B_1 + B_2 + B_3 - \frac{1}{2}(C_1 + C_2 + C_3 + C_4 \\ - C_5 - C_6)
    - \frac{1}{4}\left(D_1 + D_2 + D_3 + D_4\right) \biggr].
\end{multline}
The individual terms $A$, $B_i$, $C_i$, and $D_i$ are the same as above in Eq.~(\ref{eq:Jsum}), but with $\beta = \alpha$.

The function $K(a)$ becomes simpler under the condition $\beta = \alpha$.
Its general form (\ref{eq:Kgeneral}) simplifies to
\begin{equation}
  \label{eq:Kaaaa}
  K(a) = -\frac{\pi}{4\alpha} \left[
    \sin\left(\left|\omega - a\right| 2\alpha\right) +
    \sin\left(\left|\omega + a\right| 2\alpha\right) \right].
\end{equation}

To estimate self-product contribution to the Wigner correlation function, we should compute
\begin{equation}
  W_{aa} =
  \int\limits_{\sigma}
  J'_{aa}(\sigma_a)
  ~\mathrm{d}\sigma_a.
\end{equation}
Integration of $J'_{aa}$ here follows the very same logic as above for $J'_{ab}$.
The integral acts on the terms $A$, $B_i$, $C_i$, and $D_i$ turning the result to the form
\begin{multline}
\label{eq:Waa}
  W_{aa} =
  \frac{4}{B^2}\frac{I^2(\sigma_a)}{R_{a1}^2 R_{a2}^2}
  \biggl[ A' - B'_1 + B'_2 + B'_3 - \frac{1}{2}(C'_1 + C'_2 + C'_3 + C'_4 \\ - C'_5 - C'_6)
  - \frac{1}{4}\left(D'_1 + D'_2 + D'_3 + D'_4\right) \biggr],
\end{multline}
where terms with the primes are integrals of the original terms.
For example
\begin{equation}
  A' = \int\limits_{\sigma}
  A
  ~\mathrm{d}\sigma_a
  = \int\limits_{\sigma}
  A \,\mathrm{d} \chi \,\mathrm{d}\psi,
\end{equation}
and so on for all the terms $A$, $B_i$, $C_i$, and $D_i$.

This case is more complicated than the previous one due to the presence of cosines in the terms.
To simplify the situation, we can notice that all of the integrals can be considered as particular cases of the following integral
\begin{multline}
  M(a,b,c,d) =
  \iint\limits_{\sigma}
  \cos(2b \alpha(\chi,\psi)) \cos(2c\alpha(\chi,\psi)) \\
  \times
  \cos(2d\alpha(\chi,\psi))
  \frac{\sin (|\omega - a| 2\alpha(\chi,\psi))}{\alpha(\chi, \psi)} \,\mathrm{d} \chi \,\mathrm{d}\psi.
\end{multline}
Also, we can notice that after trigonometric transforms the integration problem can be reduced to a following integral
\begin{equation}
  \label{eq:m-general}
    m_n(a,b,c,d) =
    \iint\limits_{\sigma}
  \frac{\sin \bigl(2 f_n(a,b,c,d) \, \alpha(\chi,\psi)\bigr)}{\alpha(\chi, \psi)}
  \,\mathrm{d} \chi \,\mathrm{d}\psi,
\end{equation}
where $f_n$ denotes a remaining function, which depends only on the constants $a$, $b$, $c$, and $d$.
The original integral $M(a,b,c,d)$ can be expressed as a sum of these integrals
\begin{equation}
    \label{eq:M}
  M(a,b,c,d) =
  \frac{1}{8}
  \sum_{n=1}^8
  m_n(a,b,c,d)
  + \frac{1}{8} \sum_{n=1}^8
  m_n(-a,b,c,d).
\end{equation}
The remaining constants $f_n$ have the following form
\begin{align}
  f_1(a,b,c,d) &= |\omega - a| + d + c + b, \\
  f_2(a,b,c,d) &= |\omega - a| + d + c - b, \\
  f_3(a,b,c,d) &= |\omega - a| + d - c + b, \\
  f_4(a,b,c,d) &= |\omega - a| + d - c - b, \\
  f_5(a,b,c,d) &= |\omega - a| - d + c + b, \\
  f_6(a,b,c,d) &= |\omega - a| - d + c - b, \\
  f_7(a,b,c,d) &= |\omega - a| - d - c + b, \\
  f_8(a,b,c,d) &= |\omega - a| - d - c - b.
\end{align}

Now, we can express the individual contributions in a unified way in terms of the function $M(a,b,c,d)$ (\ref{eq:M}).
Namely
\begin{align}
  A'  &= M(0, B, \omega, \omega_0) \\
  B'_1 &= M(B,  0, \omega, \omega_0) \\
  B'_2 &= M(\omega,  0, B, \omega_0) \\
  B'_3 &= M(\omega_0, 0, B, \omega) \\
  C'_1 &= M(\omega - B,  0, 0, \omega_0) \\
  C'_2 &= M(\omega + B,  0, 0, \omega_0) \\
  C'_3 &= M(\omega_0 - B, 0, 0, \omega) \\
  C'_4 &= M(\omega_0 + B, 0, 0, \omega) \\
  C'_5 &= M(\omega_0 - \omega, 0, 0, B) \\
  C'_6 &= M(\omega_0 + \omega, 0, 0, B) \\
  D'_1 &= M(\omega_0 - \omega - B, 0, 0, 0) \\
  D'_2 &= M(\omega_0 - \omega + B, 0, 0, 0) \\
  D'_3 &= M(\omega_0 + \omega - B, 0, 0, 0) \\
  D'_4 &= M(\omega_0 + \omega + B, 0, 0, 0).
\end{align}

By plugging these equations to the original expression (\ref{eq:Waa})
we arrive to the resulting self-product contribution $W_{aa}$ over a continuous source.

Similar to the case of the cross-product contribution, only one integral $m_n(a,b,c,d)$ contains information regarding the geometry of the problem.
To obtain a particular form of the self-product contribution to the Wigner correlation, we should solve $m_n(a,b,c,d)$ either analytically or numerically.

\section{Exact solutions}
\label{section-exact-solutions}
It turns out that it is possible to obtain fully analytical solutions for the functions $L(\omega)$ (\ref{eq:L-general}) and $m_n(a,b,c,d)$ (\ref{eq:m-general}) for two classical cases of sources of rectangular and circular shapes with homogeneous distribution of intensities over their surfaces.
These solutions will provide vivid insights into properties of the Wigner correlations regarding the scales of the resulting correlations patterns.
These two cases are also interesting because they constitute classical results of the coherence theory described in~\citet{Born_Wolf} and can be used for comparison against new results obtained here.

\subsection{A rectangular source}
Let us consider a distant source of a rectangular shape of angular sizes $2\chi_0$ and $2\psi_0$ and a constant intensity $I_0$ over its surface.
The equation for $L(\omega)$ (\ref{eq:L-general}) in this case is factorized into two identical integrals
\begin{equation}
  L(\omega) = I_0^2 \int\limits_{-\chi_0}^{\chi_0} \mathrm{e}^{i2 k p\chi} \,\mathrm{d}\chi ~
  \int\limits_{-\psi_0}^{\psi_0} \mathrm{e}^{i2 k q\psi}\,\mathrm{d}\psi.
\end{equation}
These integrals resemble, up to a factor of two in the exponent, a typical integral from the diffraction theory~\citep[see][chap.~8, \S 8.5.1]{Born_Wolf}.
A solution for this integral is known
\begin{equation}
\label{eq:L-rectangle}
  L(\omega) = I_0^2 \frac{\sin(2kp\chi_0)}{kp}\,\frac{\sin(2kq\psi_0)}{kq}.
\end{equation}
This result is almost identical to the coherence pattern observed in classical interferometry due to the van Cittert–Zernike theorem, but here it is twice smaller in scale.

The integral $m_n(a,b,c,d)$ for the rectangular source can be written a form
\begin{equation}
\label{eq:m-rectangle}
  m_n =
  \int\limits_{-\chi_0}^{\chi_0}
  \int\limits_{-\psi_0}^{\psi_0}
  \frac{\sin\bigl(2f_n (p\chi + q\psi)/c\bigr)}{(p\chi + q\psi)/c}
  \,\mathrm{d}\chi
  \,\mathrm{d}\psi.
\end{equation}
Here we omit the arguments $a$, $b$, $c$, and $d$ for clarity.
An analytical solution for this integral is
\begin{equation}
  \begin{aligned}
  \label{eq:m-rectangle-result}
  m_n =
  \frac{2 c^2}{pq}\Biggl[ \frac{p \chi_0 + q \psi_0}{c} \,&\mathrm{Si}\biggl( 2 f_n \frac{p \chi_0 + q \psi_0}{c} \biggr) \\
                        - \frac{p \chi_0 - q \psi_0}{c} \,&\mathrm{Si}\biggl( 2 f_n \frac{p \chi_0 - q \psi_0}{c} \biggr) \\
                        - \frac{1}{f_n} &\sin\biggl(2f_n\frac{p\chi_0}{c}\biggr) \sin\biggl(2f_n\frac{q\psi_0}{c}\biggr) \Biggr].
  \end{aligned}
\end{equation}
Details of the computation can be found in Appendix~\ref{appendix:integral-rectangle}.

\begin{figure}[t!]
    \centering
    \includegraphics{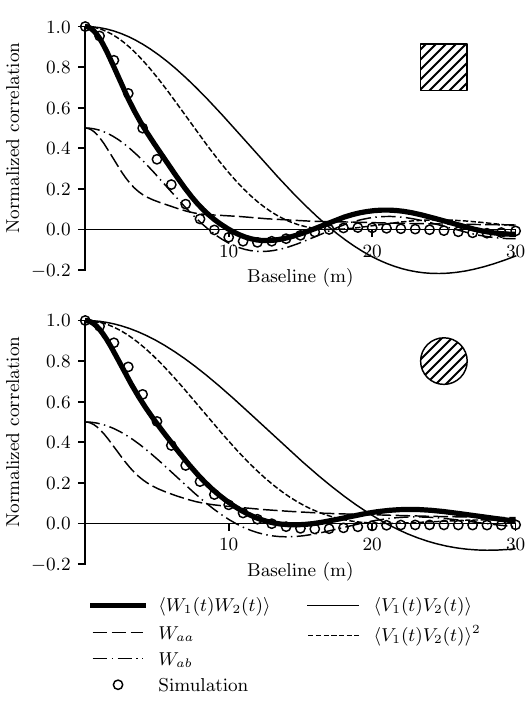}
    \caption{The Wigner correlations, $\langle W_1(t) W_2(t) \rangle$ (\ref{eq:w1w2cont}), for circular and rectangular sources of homogeneous intensity distribution ($W_{aa}$ and $W_{ab}$ are self-product and cross-product contributions correspondingly).
     The diameter of the circular source and a side of the rectangular source are $0.5^{\circ}$.
     The frequency is 2\,GHz.
     The results of the numerical simulations are also depicted on the plot.
     For comparison, the plot shows the coherence of the field fluctuations $\langle V_1(t) V_2(t) \rangle$ and its square $\langle V_1(t) V_2(t) \rangle^2 = \langle I_1(t) I_2(t) \rangle/2 - \bar{I}_1\bar{I}_2/8$, which is a part of the intensity correlations relevant to the source shape. }
    \label{fig-examples}
\end{figure}

Now the obtained solutions (\ref{eq:L-rectangle}) and (\ref{eq:m-rectangle}) can be plugged into Eqs.~(\ref{eq:Wab-final}) and (\ref{eq:M}) to obtain a particular shape of the Wigner coherence pattern.
Figure~\ref{fig-examples} depicts a cross section of resulting pattern along the $x$-axis of the observation plane $\sigma$ together with the classical results from the van Cittert-Zernike theorem.

\subsection{A circular source}
To solve the problem for a source of a circular shape, we can exploit its axial symmetry.
We follow usual practice and represent the coordinates of both the point on the source and the observation points in polar coordinates.
Namely
\begin{align}
  \chi &= \rho \cos \theta, \\
  \psi &= \rho \sin \theta
\end{align}
for the point on the source, and
\begin{align}
  p &= w \cos \eta, \\
  q &= w \sin \eta
\end{align}
for the relative coordinate of the observation point.

The equation $L(\omega)$ becomes
\begin{equation}
  L(\omega) =
  I_0^2
  \int\limits_0^a\int\limits_0^{2\pi}
  \mathrm{e}^{i2k w \rho \cos{(\theta - \eta)}}
  \rho
  \,\mathrm{d}\rho
  \,\mathrm{d}\theta,
\end{equation}
where $a$ denotes an angular radius of the source.
The solution for this integral is known from the classical theory of diffraction~\citep[see][chap.~8, \S 8.5.2]{Born_Wolf}
\begin{equation}
    L(\omega) = \pi a^2 \frac{2J_1(\kappa)}{\kappa}\mathrm{e}^{i \phi}.
\end{equation}
The constants $\kappa$ and $\phi$ are defined as follows
\begin{align}
    \kappa &= 2 k a \sqrt{p^2 + q^2} = 2 k a w, \\
    \phi   &= \frac{2\pi}{\lambda} \zeta.
\end{align}
Thus, the coherence pattern associated with this contribution is identical to the classical interferometry pattern~\citep[see][chap.~8, \S 8.5.2]{Born_Wolf}, but twice smaller in scale.

In the integral appearing in $m_n(a,b,c,d)$ (\ref{eq:m-general}) has a following form in the introduced polar coordinates
\begin{equation}
  \label{eq:J4}
  m_n =
  \int\limits_0^a\int\limits_0^{2\pi}
  \frac{\sin\bigl(2/c\, f_n w \rho \cos(\theta - \eta)\bigr)}
       {     1/c\, w \rho \cos(\theta - \eta) }
  \rho
  \,\mathrm{d}\rho
  \,\mathrm{d}\theta.
\end{equation}
Here once again we omit the arguments $a$, $b$, $c$, and $d$ for clarity.
It is possible to solve this integral analytically.
A solution, which one can obtain is
\begin{equation}
  \label{eq:J4result}
  m_n = \frac{2\pi c \rho}{w}
  \Bigl[
  \mu J_0(\mu) - J_1(\mu)
  + \frac{\pi\mu}{2} \bigl[
  J_1(\mu) \mathbf{H}_0(\mu) -
  J_0(\mu) \mathbf{H}_1(\mu)
  \bigr]
  \Bigr],
\end{equation}
where $J_{\nu}(\mu)$ and $\mathbf{H}_{\nu}(\mu)$ are corresponding Bessel and Struve functions of first kind and of zero and first order with the following arguments
\begin{equation}
  \mu = 2\frac{f_n}{c} w \rho.
\end{equation}

Computation of Eq.~(\ref{eq:J4result}) requires of a number of transforms.
All details can be found in Appendix~\ref{appendix:integral}.

The obtained solutions (\ref{eq:L-rectangle}) and (\ref{eq:m-rectangle}) can be used now to compute the Wigner coherence pattern for a circular source.
Figure~\ref{fig-examples} illustrates a cross section of a resulting pattern for a particular source size together with the results for the established techniques.

For both of the examples considered here we see that the scale of the Wigner coherence function appears about twice smaller in comparison to the shape of the classical interferometric pattern.
This effect appears to be a vivid difference between the Wigner and classical interferometry based on the correlation of the fields.

\section{Numerical simulations}
\label{section-simulation}
From the analytical computations presented above, we can see that the resulting pattern for the Wigner correlation is about twice smaller in comparison to the correlations of the field fluctuations or their intensities.
This effect can be also observed in numerical simulations, which independently supports the analytical findings.
In fact, this is how the effect was observed for the first time.

Since the purpose of this work is to introduce a new technique, the goal here is to perform only conceptual proof-of-principal simulations rather then detailed simulations corresponding to a very particular problem.
Such a conceptual approach allow us to focus on the main effect related to the scale of the coherence pattern and avoid consideration of multiple other effects, which are not relevant for the time being.
The key elements of these simulations are:
\begin{enumerate}
  \item simulation of incoming radiation from a distance source in form of discrete-time series representing a single polarization
  \item computation of the Wigner function for the time series
  \item estimation of the Wigner correlation at a given frequency from the functions obtained on the previous step
\end{enumerate}

The source code for these simulations is openly available.\footnote{\url{https://github.com/vlenok/wigner-interferometry}}
The results of the simulations performed with the code for homogeneous circular and rectangular sources are depicted on Figure~\ref{fig-examples} alongside with the theoretical results.

Sections below highlight few aspects related to the methods used in the simulation code.
Consult the code for particular details of implementation.

\subsection{Incoming fields}
To compute fields arriving from a distant source, we represent the source as a point cloud with a grid alignment of the individual points.
The grid is defined in terms of angular distances $\chi$ and $\psi$ visible from the origin of the observation plane $\sigma$ ($\chi \approx x/R$, $\psi \approx y/R$, see Figure~\ref{fig:geometry}).
Each point is considered as an independent source of Gaussian white noise (with $\omega_0=0$, where $\omega_0$ is defined in Eq.~(\ref{eq:power-spectrum})) making this discrete model of a source noncoherent by construction.
The lower cut off of the spectrum is not used here to avoid numerical limitations related to aliasing of the Wigner time-frequency distribution.

A sample of the Gaussian noise obtained for every point of the source propagates a slightly different distance depending on position of the observation point at the plane $\sigma$.
To account for this difference, a corresponding shift in the time domain is introduced and computed for every location of the radiation and observation points individually.
To perform this shift in the generated time series, the code uses a discrete-time phase-shift filter with finite impulse response filter designed with the window method~\citep{phase-shifter}.
Due to its design, this filter naturally introduces a high-frequency cut off on the level of 80\% from the highest frequency governed by the Nyquist sampling theorem.
To obtain a resulting time series for the field at a given observation point, all the time series from the individual points of the source are summed up.

\subsection{The Wigner time-frequency distribution}
There are multiple known algorithms enabling computation of Wigner functions from for time series.
For this simulation, a variation of the algorithm described in \citet{chassande-mottin} is used since it is less susceptible to numerical artifacts related to the aliasing effects common for the quadratic time-frequency distributions (see for example \citet{flandrin-geometry}).

\subsection{The Wigner correlation}
To estimate the Wigner correlation, the code selects temporal variations of the Wigner distributions at a given frequency and multiplies them for different baselines.
Then, the resulting time series of the products are summed up at every baseline to obtain an unnormalized estimator of the Wigner correlation.
To normalize this estimator, it is divided over the value at the zero baseline.

\section{Discussion}
In this paper we considered Wigner interferometry, a new method of interferometric measurements based on a correlation of Wigner functions constructed from the fields of a distant source.
The computations above revealed that this correlation has properties, which differ from those known for the correlation of the fields themselves.
Namely, in comparison to the correlation of the fields, the Wigner correlation contains twice higher frequencies at a given baseline and has about twice smaller scale of the resulting correlation pattern\footnote{
  A similar effect was observed numerically for the Unterberger time-frequency distribution.
  It may indicate that the effects discussed in this work have a more general nature and manifest themselves in other quadratic time-frequency distributions.
}.

A natural question arises whether the obtained findings violate the known results.
To address this question, we should notice what kind of physical quantities are fundamental for one or the other types of interferometry.
The central quantity for classical radio interferometry and intensity interferometry is the second-order correlation of the field vibrations (see Eqs.~\ref{eq-intro-vcz} and \ref{eq-intro-II}).
The only difference appearing in intensity interferometry is that this quantity is squared.
In contrast, the central quantity in Wigner interferometry is a fourth-order correlation function of the field vibrations (Eq.~\ref{eq-gamma}), which naturally appears as a result of Wigner correlation.
Thus, the known results regarding the diffraction limit and the van Cittert-Zernike theorem are related to the second-order correlation, and the new results describe behavior of a different type of the correlation.

The computations above revealed that the Wigner correlation is a sum of the cross-product and self-product contributions (Eq.~\ref{eq:w1w2cont}).
It seems possible to assign a certain physical interpretation to these contributions due to their connection to particular locations on the source.
We can interpret them as interference between the correlation functions originating from the corresponding locations.
Namely, the cross-product contribution reflects an interference between the correlation functions originating from the axially symmetric points on the source, and the self-product contribution reflects an interference of the correlation function with itself.

The spatial properties of the Wigner correlation open a fundamental opportunity to improve the angular resolution of interferometric measurements by a factor of two.
To achieve this improvement, it is necessary to develop a method enabling image reconstruction from the measured Wigner correlations.
Even though, the present work does not provide it, it does not seem that there are fundamental obstacles in such a development.

In conclusion, we can note that even though Wigner interferometry is considered in this work in the context of radio astronomy, the obtained theoretical results likely have a more general nature.
The reason for this possibility is the fact that the starting point of the computations are generic spherical waves (Eq.~\ref{eq:singleV}) without any intrinsic assumptions regarding the nature of the fields.
Also, given the history of the van Cittert-Zernike theorem and the intensity interferometry, called also as the Hanbury Brown and Twiss effect, it is reasonable to expect that the principles of Wigner interferometry are applicable to quantum systems even though the obtained results is purely classical.

\section{Conclusions}
Wigner interferometry presented in this work opens a new fundamental opportunity to improve angular resolution of interferometric measurements.
This technique can have different practical applications and imaging of the black hole shadows seems to be among the obvious ones.
Such observations are profoundly important for studies of black hole physics, general relativity, jet formation mechanism, etc.~\citep{boccardi-review, astro2020}.
However, already now global-scale instruments like the Event Horizon Telescope are required to make possible imaging of such small objects and higher frequencies or space baselines are required to observe their finer details~\citep{ngeht, bhex}.
However, when Wigner interferometry is turned into a practical technique, it has a potential to push the angular resolution achievable by any given interferometer by an additional factor of two.

\begin{acknowledgements}
  The author would like to thank Dominik J.~Schwarz and Dmitriy~Kostunin for discussing the matters described in this paper and for helping with its preparation and Travis~Dore for proofreading the manuscript.
  The author would like to specially thank Vladimir~Sotnikov for his invaluable assistance in preparing the software for public release.
  This work was funded by the Deutsche Forschungsgemeinschaft (DFG, German Research Foundation) – project number 460248186 (PUNCH4NFDI).
\end{acknowledgements}

%
  \bibliographystyle{aa} 
  \bibliography{ref} 
\begin{appendix} 




\section{Coherence of field fluctuations}
\label{appendix-VV}
In this appendix we will reproduce the main statement of the van Cittert-Zernike theorem in the notation used in this work.
We follow the logic described in~\citet[chap.~10, \S 10.4.2]{Born_Wolf}.

Let us consider only the field from the point $a$ arriving to the points $P_1$ and $P_2$ of the detection plane
\begin{align}
  V_{a1} &= \frac{1}{R_{a1}}
  A_a \left(t - R_{a1}/c\right)\,
  \cos \left[
    \omega \left( t - R_{a1}/c \right)
    \right],\\
  V_{a2} &= \frac{1}{R_{a2}}
  A_a \left(t - R_{a2}/c\right)\,
  \cos \left[
    \omega \left( t - R_{a2}/c \right)
    \right].
\end{align}
If we consider their correlation,
we obtain
\begin{multline}
  \langle V_{a1}(t) V_{a2}(t) \rangle =
  \frac{1}{R_{a1}}\frac{1}{R_{a2}}
  \langle A_a \left(t - R_{a1}/c\right)
  A_a \left(t - R_{a2}/c\right) \rangle \\
  \times
  \cos\left[ \omega \frac{R_{a1} - R_{a2}}{c} \right].
\end{multline}
To average the cosine functions, we expressed them in an exponential form, which we can denote as $z_1$ and $z_2$, and used an expression $\mathrm{Re}\{z z^*\}$~\citep[see][chap.~6, \S 48]{landau2}.
This led to the resulting cosine factor.

Now we can discard retardation in the remaining correlation and notice that then it corresponds to a mean intensity of an element $\sigma_a$ centered at the point $a$ of the source
\begin{equation}
  \langle A_a \left(t - R_{a1}/c\right)
  A_a \left(t - R_{a2}/c\right) \rangle \approx
  \langle A_a(t)
  A_a(t) \rangle = \bar{I}(\sigma_a).
\end{equation}
With this simplification, the original equation simplifies to
\begin{equation}
  \langle V_{a1}(t) V_{a2}(t) \rangle =
  \frac{\bar{I}_a}{R_{a1}R_{a2}} \cos\left[\omega \frac{R_{a1} - R_{a2}}{c}\right].
\end{equation}
If we generalize this equation to an extended continuous source, we arrive at
\begin{equation}
  \langle V_{1}(t) V_{2}(t) \rangle =
  \int\limits_{\sigma}
  \frac{\bar{I}(\sigma_a)}{R_{a1}R_{a2}} \cos\left[\omega \frac{R_{a1} - R_{a2}}{c}\right]
  \,\mathrm{d}\sigma_a,
\end{equation}
or, equivalently
\begin{equation}
  \label{eq-vcz-final}
  \langle V_{1}(t) V_{2}(t) \rangle =
  \mathrm{Re}
  \int\limits_{\sigma}
  \bar{I}(\sigma_a)
  \frac{\mathrm{e}^{i k \left(R_{a1} - R_{a2}\right)}}
       {R_{a1}R_{a2}}
  \,\mathrm{d}\sigma_a,
\end{equation}
where $k = \omega/c$ denotes here the wave number and $\sigma_a$ is an element of the source surface $\sigma$ (see Figure~\ref{fig:geometry}).
This integral reproduces the integral appearing in the van Cittert-Zernike theorem.
To reproduce the statement of this theorem and vividly see, that the resulting spatial correlation pattern is related to the Fourier transform, we should use explicit coordinates for all points and use usual large distance approximations.
In particular, the following approximations for the product
\begin{equation}
  R_{a1} R_{a2} \approx R^2
\end{equation}
and the difference
\begin{multline}
  \label{eq:R-difference}
  R_{a1} - R_{a2} = \frac{1}{2R}\left[ (X_1^2 + Y_1^2) - (X_2^2 + Y_2^2) \right] \\
  - \frac{1}{R}\left[ x_a(X_1 - X_2) + y_a(Y_1 - Y_2) \right].
\end{multline}
Only the second bracket in (\ref{eq:R-difference}) is relevant the integral.
The first bracket only introduces a constant phase offset.
Also, let us measure the distances along the source in angular units instead of linear ones
\begin{align}
    \label{eq:chi-definition}
    \chi &\approx x_a/R, \\
    \label{eq:psi-definition}
    \psi &\approx y_a/R
\end{align}
and denote the coordinate differences between the observation points as
\begin{align}
  p &= X_1 - X_2, \\
  q &= Y_1 - Y_2.
\end{align}

After all these transforms, the equation for distance difference (\ref{eq:R-difference}) reduces to
\begin{equation}
  R_{a1} - R_{a2} = \zeta - (p\chi + q\psi),
\end{equation}
where
\begin{equation}
  \zeta = \frac{1}{2R}\left[ (X_1^2 + Y_1^2) - (X_2^2 + Y_2^2) \right]
\end{equation}
is a phase offset, which can be put to zero for large distances $R$.

After all these large-distance approximations, the integral (\ref{eq-vcz-final}) has a following form
\begin{equation}
  \label{eq:appendix-vcz}
  \langle V_{1}(t) V_{2}(t) \rangle =
  \iint\limits_{\sigma} I(\chi, \psi) \mathrm{e}^{-ik(p\chi + q\psi)} \,\mathrm{d}\chi\,\mathrm{d}\psi\,
\end{equation}
from which we immediately see, that the spatial coherence function is a Fourier transform of the source intensity distribution.
Now we can see that the variables $p$ and $q$ are spatial frequencies.

The equation (\ref{eq:appendix-vcz}) constitutes the statement of the van Cittert-Zernike theorem~\citep[see][chap.~10]{Born_Wolf}.

\section{Coherence of field intensities}
\label{appendix-II}
In this appendix, we will reproduce the main result of intensity interferometry.
We use the same approach as in the main text of the paper for derivation of the equations for Wigner interferometry.
Namely, first, we consider only two points, and then we generalize the obtained equation to a case of the continuous source.

Let us consider a correlation of field intensities at two points for the case of two points that only radiate as on the scheme in Figure~\ref{fig:geometry}
\begin{multline}
  \langle I_1(t) I_2(t+\tau) \rangle =
  \langle V_1(t) V_1(t) V_2(t + \tau) V_2 (t+\tau) \rangle \\
  =
  \langle
  \left[ V_{a1}(t) + V_{b1}(t) \right]
  \left[ V_{a1}(t) + V_{b1}(t) \right] \\
  \times
  \left[ V_{a2}(t + \tau) + V_{b2}(t + \tau) \right]
  \left[ V_{a2}(t + \tau) + V_{b2}(t + \tau) \right]
  \rangle.
\end{multline}
In a shorthand notation similar to the one used in the main text this correlation function becomes
\begin{multline}
  \langle I_1(t) I_2(t+\tau) \rangle =
  \langle
  \left[ a_1(0) + b_1(0) \right]
  \left[ a_1(0) + b_1(0) \right]\\
  \times
  \left[ a_2(\tau) + b_2(\tau) \right]
  \left[ a_2(\tau) + b_2(\tau) \right]
  \rangle = J_{12}(\tau)
\end{multline}
If we expand the brackets, use the Isserlis's theorem~\citep{isserlis-theorem} assuming the Gaussian nature of the radiation, and put all the terms of a type $\langle a b \rangle$ to zero, we arrive at
\begin{equation}
  \begin{split}
    J_{12}(\tau)
    &= \langle a_1(0) a_1(0) \rangle \langle a_2(\tau) a_2(\tau) \rangle
    + \langle a_1(0) a_2(\tau) \rangle \langle a_1(0) a_2(\tau) \rangle \\
    &+ \langle a_1(0) a_1(\tau) \rangle \langle a_1(0) a_2(\tau) \rangle
    + \langle b_1(0) b_1(0) \rangle \langle b_2(\tau) b_2(\tau) \rangle \\
    &+ \langle b_1(0) b_2(\tau) \rangle \langle b_1(0) b_2(0) \rangle
    + \langle b_1(0) b_2(\tau) \rangle \langle b_1(0) b_2(\tau) \rangle \\
    &+ \langle a_1(0) a_1(0) \rangle \langle b_2(\tau) b_2(\tau) \rangle
    + \langle b_1(0) b_1(0) \rangle \langle a_2(\tau) a_2(\tau) \rangle \\
    &+ 4 \langle a_1(0) a_2(\tau) \rangle \langle b_1(0) b_2(\tau) \rangle.
  \end{split}
\end{equation}
To simplify the notation further, let us use the following notation
  $\Gamma_{11}^{(a)}(0) = \langle a_1(0) a_1(0) \rangle = \langle a_1(\tau) a_1(\tau) \rangle$,
  $\Gamma_{12}^{(a)}(0) = \langle a_1(0) a_2(\tau) \rangle$, etc. for the appearing correlation functions.
Thus, we can simplify the equation above, and after collecting of term, we obtain
\begin{equation}
  \begin{split}
  J_{12}(\tau) &=  \Gamma_{11}^{(a)}(0) \Gamma_{22}^{(a)}(0) + \Gamma_{11}^{(b)}(0) \Gamma_{22}^{(b)}(0) \\
  &+ \Gamma_{11}^{(a)}(0) \Gamma_{22}^{(b)}(0) + \Gamma_{11}^{(b)}(0) \Gamma_{22}^{(a)}(0) \\
  &+ 2 \Gamma_{12}^{(a)}(\tau) \Gamma_{12}^{(a)}(\tau) + 2 \Gamma_{12}^{(b)}(\tau) \Gamma_{12}^{(b)}(\tau)
  + 4 \Gamma_{12}^{(a)}(\tau) \Gamma_{12}^{(b)}(\tau).
  \end{split}
\end{equation}
This equation has the following structure
\begin{multline}
  J_{12}(\tau) = \left( \Gamma_{11}^{(a)}(0) + \Gamma_{11}^{(b)}(0) \right)
  \left( \Gamma_{22}^{(a)}(0) + \Gamma_{22}^{(b)}(0) \right) \\
  + 2\left(\Gamma_{12}^{(a)}(\tau) + \Gamma_{12}^{(b)}(\tau) \right)^2.
\end{multline}
Since the correlations of a type $\Gamma_{11}^{(a)}(0)$ by definition are averaged intensities of the field
\begin{equation}
  \Gamma_{11}^{(a)}(0) = \frac{1}{2}\bar{I}_1^{(a)},
\end{equation}
the structure of the equation is such that the first bracket is a contribution to the intensity at the point $P_1$ from the two points of the source.
The second bracket is the same, but for the point $P_2$.
The last bracket is the correlation function of the two-point source visible from the points $P_1$ and $P_2$.

We can generalize this result to a case of continuous source.
If we come back to the original notation for the correlation, we arrive to a generalized form
\begin{equation}
  \label{eq-intensity-interferometry-main}
  \langle I_1(t) I_2(t+\tau) \rangle = \frac{1}{4} \bar{I}_1 \bar{I}_2 + 2 \left[\langle V_1(t) V_2(t + \tau) \rangle\right]^2,
\end{equation}
where the averaged intensity of the field visible at the points $P_1$ and $P_2$, and the correlation function are integrated over the entire area of the source
\begin{align}
  \bar{I}_{1,2}     &= \int\limits_{\sigma} \bar{I}_{1,2}(\sigma_a) \,\mathrm{d}\sigma_a, \\
  \langle V_1(t) V_2(t + \tau) \rangle &= \int\limits_{\sigma} \langle V_{\vec{a}1}(t) V_{\vec{a}2}(t + \tau) \rangle \,\mathrm{d}\sigma_a.
\end{align}

The generalized equation (\ref{eq-intensity-interferometry-main}) reproduces the main equation of the intensity interferometry~\citep[see][Eq.~(3.36)]{hanbury-brown}.
If we assume $\tau = 0$, which is the case for the intensity interferometry, the correlation function $\langle V_1(t) V_2(t + \tau) \rangle$ in~(\ref{eq-intensity-interferometry-main}) is identical to the correlation function~(\ref{eq-vcz-final}).

\section{The integral $m_n$ for a rectangular source}
\label{appendix:integral-rectangle}
For completeness, here we compute the integral $m_n(a,b,c,d)$ for the case of a rectangular source explicitly.

First, it seems reasonable to obtain sine integral function in one of the integrals in Eq.~(\ref{eq:m-general}).
We transform the integral over $\chi$ to
\begin{equation}
  I = \frac{1}{p'}
  \int\limits_{-\chi_0}^{\chi_0}
  \frac{\sin\bigl(2f_n (p'\chi + q'\psi)\bigr)}{(p'\chi + q'\psi)}
  \,\mathrm{d}\left(2f_n (p'\chi + q'\psi)\right),
\end{equation}
where $p'=p/c$ and $q'=q/c$.

Then to obtain the standard definition of the sine integral, we split the integral into two and flip the limits in one of them
\begin{equation}
    I = \frac{1}{p'} \left[ -\int\limits_0^{-\chi_0}\dots + \int\limits_0^{\chi_0}\dots \right].
\end{equation}
This gives a solution of the first integral in Eq.~(\ref{eq:m-general})
\begin{equation}
\label{eq:m-square-inter}
  I = \frac{1}{p'}\left[
  - \mathrm{Si}\left(2f_n(-p'\chi_0 + q'\psi)\right)
  + \mathrm{Si}\left(2f_n(p'\chi_0 + q'\psi)\right)
  \right].
\end{equation}

Let us turn our attention now to the second integral (over $\psi$).
It acts to the right-hand side of Eq.~(\ref{eq:m-square-inter}) and result into two similar integrals
\begin{multline}
\label{eq:J}
    J =
    -\frac{1}{p'}\int\limits_{-\psi_0}^{\psi_0}
    \mathrm{Si}\left(2f_n(-p'\chi_0 + q'\psi)\right)
    \,\mathrm{d} \psi \\
    +\frac{1}{p'}\int\limits_{\psi_0}^{\psi_0}
    \mathrm{Si}\left(2f_n(p'\chi_0 + q'\psi)\right)
    \,\mathrm{d} \psi = \frac{- J_1 + J_2}{p'}.
\end{multline}

Let us consider one of them
\begin{equation}
\label{eq:J2}
    J_2 =
    \int\limits_{-\psi_0}^{\psi_0}
    \mathrm{Si}\left(2f_n(p'\chi_0 + q'\psi)\right)
    \,\mathrm{d} \psi.
\end{equation}
We can see that this integral is essentially an integral of sine integral, which can be solved by integration by parts
\begin{equation}
\label{eq:I3interm}
    y = \int \mathrm{Si}(ax)\,\mathrm{d}x =
    x \,\mathrm{Si} - \int x\,\left[ \frac{\mathrm{d}}{\mathrm{d}x} \mathrm{Si}(ax)\right]\,\mathrm{d}x.
\end{equation}
The remaining derivative is
\begin{equation}
    \frac{\mathrm{d}}{\mathrm{d}x} \mathrm{Si}(ax) =
     \frac{\mathrm{d}}{\mathrm{d}x} \int\limits_0^{ax}\frac{\sin t}{t}\,\mathrm{d}t = \frac{\sin ax}{x}.
\end{equation}
With this derivative Eq.~(\ref{eq:I3interm}) becomes
\begin{equation}
    y =
    x \,\mathrm{Si} + \frac{1}{a}\cos ax.
\end{equation}
We can use this result to solve the integral (\ref{eq:J2})
\begin{multline}
    J_2 =
    \frac{1}{q'} \biggl[
     A\,\mathrm{Si}\left(2f_nA\right)
     + \frac{1}{2f_n} \cos\left(2f_nA\right) \\
    -  B\,\mathrm{Si}\left(2f_nB\right) 
    - \frac{1}{2f_n} \cos\left(2f_nB\right)
    \biggr],
\end{multline}
where
\begin{align}
    A &= p'\chi_0 + q' \psi_0, \\
    B &= p'\chi_0 - q' \psi_0.
\end{align}
The solution for the integral $J_1$ in Eq.~(\ref{eq:J}) can be obtained by substituting $p'$ with $-p'$ in the solution for $J_2$.
If we combine now the solutions for the integrals $J_1$ and $J_2$ and simplify the resulting equation, we obtain
\begin{multline}
\label{eq:Jresult}
    I =
    \frac{2}{p'q'} \biggl[
     A\,\mathrm{Si}\left(2f_nA\right)
    - B\,\mathrm{Si}\left(2f_nB\right) \\
    + \frac{1}{2f_n} \left[ \cos\left(2f_nA\right) - \cos\left(2f_nB\right) \right]
    \biggr].
\end{multline}
The cosines can be further simplified to
\begin{multline}
\label{eq:cosine-simplification}
    \cos\left(2f_n ( p' \chi_0 + q' \psi_0)\right) -
    \cos\left(2f_n ( p' \chi_0 - q' \psi_0)\right) \\
    = -2\sin \left(2f_n p'\chi_0\right) \sin\left(2f_nq'\psi_0\right).
\end{multline}
A combination of Eqs.~(\ref{eq:Jresult}) and (\ref{eq:cosine-simplification}) results into Eq.~(\ref{eq:m-rectangle-result}).

\section{The integral $m_n$ for a circular source}
\label{appendix:integral}
For completeness, here we compute the integral $m_n(a,b,c,d)$ for a circular source.

Let us first consider the integral
\begin{equation}
  \label{eq:J4p}
  I = \int\limits_0^{2\pi}
  \frac{\sin\bigl(2/c\, f_n w \rho \cos(\theta - \eta)\bigr)}
       {     1/c\, w \rho \cos(\theta - \eta) }
  \,\mathrm{d}\theta,
\end{equation}
which is the internal integral in Eq.~(\ref{eq:J4}).
To solve it, we multiply the integral to $2f_n$ to obtain the denominator the same as the argument of the sine, which we temporarily denote as $x$ here ($x = 2/c\, f_n w \rho \cos(\theta - \eta)$), and then we expand the sine in a Taylor series
\begin{equation}
  I =
  2 f_n
  \int\limits_0^{2\pi}
    \,\mathrm{d}\theta\,
  \frac{1}
       { x }
       \left[ x - \frac{x^3}{3!} + \frac{x^5}{5!} - \dots \right]
\end{equation}
from which we obtain
\begin{equation}
  I =
  2 f_n
  \int\limits_0^{2\pi}
    \,\mathrm{d}\theta\,
    \sum_{n=0}^{\infty}
    \frac{(-1)^n}{(2n+1)!}\, x^{2n}.
\end{equation}
Now we can swap the integral and the sum
\begin{equation}
  I =
  2 f_n
  \sum_{n=0}^{\infty}
  \frac{(-1)^n}{(2n+1)!}
  \left[\frac{2}{c}f_n w \rho \right]^{2n}
  \int\limits_0^{2\pi}
  \,\mathrm{d}\theta\,
  \cos^{2n} (\theta - \eta).
\end{equation}
The remaining integral of the cosine is known~\citep[see][2.5.12.44]{prudnikov1}.
If we use this known solution, $I$ becomes
\begin{equation}
  I =
  2 f_n
  \sum_{n=0}^{\infty}
  \frac{(-1)^n}{(2n+1)!}
  \left[\frac{2}{c}f_n w \rho \right]^{2n}
  \frac{2\pi}{2^{2n}}
  \frac{(2n)!}{(n!)^2}.
\end{equation}
For convenience we can denote
\begin{equation}
  \mu = 2 \frac{f_n}{c} w \rho
\end{equation}
and simplify the equation, this gives a following expression
\begin{equation}
  \label{eq:J4interm}
  I =
  \frac{4\pi f_n}{\mu}
  \sum_{n=0}^{\infty}
  \frac{(-1)^n}{2^{2n}\, n!\, n!}
  \frac{\mu^{2n+1}}{2n+1}.
\end{equation}
The later fraction can be expressed in an integral form
\begin{equation}
  \frac{\mu^{2n+1}}{2n+1} =
  \int\limits_0^{\mu} t^{2n}\,\mathrm{d}t.
\end{equation}
If we use this form in Eq.~(\ref{eq:J4interm}) and rearrange it, we arrive at
\begin{equation}
  I =
  \frac{4\pi f_n}{\mu}
  \int\limits_0^{\mu}
  \mathrm{d}t\,
  \sum_{n=0}^{\infty}
  \frac{(-1)^n}{n!n!} \left(\frac{t}{2} \right)^{2n},
\end{equation}
where the remaining sum is known to be a Bessel function~\citep[see][5.2.10.1]{prudnikov1}. This brings us to an intermediate result
\begin{equation}
  I =
  \frac{4\pi f_n}{\mu}
  \int\limits_0^{\mu}
  J_0(t)\,
  \mathrm{d}t.
\end{equation}

By using the obtained intermediate result, we can write the original integral for $m_n$ in a form
\begin{equation}
  \label{eq:J4doubleint}
  m_n =
  \frac{2\pi c}{w}
  \int\limits_0^a
  \int\limits_0^{\mu(\rho)}
  J_0(t)
  \,\mathrm{d} t
  \,\mathrm{d} \rho.
\end{equation}

It is known that an integral of a Bessel function can be expressed as a series of Bessel functions~\citep[see][1.8.1.9]{prudnikov2}
\begin{equation}
  \label{eq:BesselIntegral}
  \int\limits_0^y J_0(x) \,\mathrm{d} x = 2 \sum_{k=0}^{\infty} J_{2k+1} (y).
\end{equation}
Let us use this representation to transform Eq.~(\ref{eq:J4doubleint}).
First, the internal integral over $t$ becomes a sum
\begin{equation}
  m_n = 2 \frac{2\pi c}{w} \int\limits_0^a \mathrm{d} \rho \, \sum_{k=0}^{\infty} J_{2k+1} (\mu).
\end{equation}
Now we can swap the integral and the sum
\begin{equation}
  m_n = 2 \frac{2\pi c}{w}
  \sum_{k=0}^{\infty}
  \int\limits_0^a J_{2k+1} (\mu) \, \mathrm{d} \rho
\end{equation}
and use the same representation (\ref{eq:BesselIntegral}) again to arrive to a following double sum
\begin{equation}
  m_n =
  \frac{4\pi c^2}{f_n w^2} \sum_{k=0}^{\infty} \sum_{l=0}^{\infty} J_{2(k+l) + 2} (\mu),
\end{equation}
which due to repetition of the Bessel functions can be expressed as a single sum
\begin{equation}
  m_n =
  \frac{4\pi c^2}{f_n w^2} \sum_{k=1}^{\infty} k J_{2k} (\mu).
\end{equation}
The remaining sum is known~\citep[see][5.7.2.10]{prudnikov2} and can be expressed via the Bessel and Struve functions of the first kind and of the zero and first order.
This brings us to the final result (\ref{eq:J4result}).

\end{appendix}
\end{document}

%% file: tikzit/styles.tikzstyles
\tikzstyle{black dot}=[fill=black, draw=black, shape=circle, minimum size=1mm, inner sep=0pt, outer sep=0pt]
\tikzstyle{nice head}=[{-{Stealth[length=2.5mm, width=1mm]}}, semithick]

\tikzstyle{dashed line}=[-, dashed, semithick, dash pattern=on 4mm off 2mm]
\tikzstyle{thick line}=[-, semithick]
\tikzstyle{semithick line}=[-]

%% file: ref.bib
@book{prudnikov1,
  title     = "Integrals and series",
  author    = "Prudnikov, A. P. and  Bry\u{c}kov, J. A. and  Mari\u{c}ev, O. I.",
  year      =  1998 ,
  publisher = "Gordon \& Breach",
  address   = "New York",
  volume    = 1
}

@book{prudnikov2,
  title     = "Integrals and series",
  author    = "Prudnikov, A. P. and  Bry\u{c}kov, J. A. and  Mari\u{c}ev, O. I.",
  year      =  1998 ,
  publisher = "Gordon \& Breach",
  address   = "New York",
  volume    = 2
}

@book{Born_Wolf,
  place={Cambridge},
  edition={7},
  title={Principles of Optics},
  publisher={Cambridge University Press},
  author={Born, Max and Wolf, Emil},
  year={2019}
}

@book{landau2,
title = "The Classical Theory of Fields",
publisher = "Pergamon",
edition = 4,
address = "Amsterdam",
year = 1975,
volume = 2,
series = "Course of Theoretical Physics",
author = "Landau, L. D. and Lifshitz, E. M."
}

@book{hanbury-brown,
  place={London},
  title={The intensity interferometer. Its application to astronomy},
  publisher={Taylor \& Francis Ltd},
  author={Hanbury Brown, R.},
  year={1974}
}

@BOOK{thompson2017,
       author = {{Thompson}, A. Richard and {Moran}, James M. and {Swenson}, Jr., George W.},
        title = "{Interferometry and Synthesis in Radio Astronomy, 3rd Edition}",
         year = 2017,
         place = "Cham",
         publisher = "Springer",
}

@ARTICLE{optical-interferometry,
       author = {{Eisenhauer}, Frank and {Monnier}, John D. and {Pfuhl}, Oliver},
        title = "{Advances in Optical/Infrared Interferometry}",
      journal = {\araa},
         year = 2023,
        month = aug,
       volume = {61},
        pages = {237-285},
          doi = {10.1146/annurev-astro-121622-045019},
archivePrefix = {arXiv},
       eprint = {2303.00453},
 primaryClass = {astro-ph.IM},
       adsurl = {https://ui.adsabs.harvard.edu/abs/2023ARA&A..61..237E}
}

@article{vla-main,
doi = {10.1088/2041-8205/739/1/L1},
url = {https://doi.org/10.1088/2041-8205/739/1/L1},
year = {2011},
month = {aug},
publisher = {The American Astronomical Society},
volume = {739},
number = {1},
pages = {L1},
author = {Perley, R. A. and Chandler, C. J. and Butler, B. J. and Wrobel, J. M.},
title = {THE EXPANDED VERY LARGE ARRAY: A NEW TELESCOPE FOR NEW SCIENCE},
journal = {The Astrophysical Journal Letters}
}

@INPROCEEDINGS{meerkat-main,
       author = {{Jonas}, J. and {MeerKAT Team}},
        title = "{The MeerKAT Radio Telescope}",
    booktitle = {MeerKAT Science: On the Pathway to the SKA},
         year = 2016,
        month = jan,
          eid = {1},
        pages = {1},
          doi = {10.22323/1.277.0001},
       adsurl = {https://ui.adsabs.harvard.edu/abs/2016mks..confE...1J}
}

@ARTICLE{lofar-main,
       author = {{van Haarlem}, M.~P. and {Wise}, M.~W. and {Gunst}, A.~W. and {Heald}, G. and {McKean}, J.~P. and {Hessels}, J.~W.~T. and {de Bruyn}, A.~G. and {Nijboer}, R. and {Swinbank}, J. and {Fallows}, R. and {Brentjens}, M. and {Nelles}, A. and {Beck}, R. and {Falcke}, H. and {Fender}, R. and {H{\"o}randel}, J. and {Koopmans}, L.~V.~E. and {Mann}, G. and {Miley}, G. and {R{\"o}ttgering}, H. and {Stappers}, B.~W. and {Wijers}, R.~A.~M.~J. and {Zaroubi}, S. and {van den Akker}, M. and {Alexov}, A. and {Anderson}, J. and {Anderson}, K. and {van Ardenne}, A. and {Arts}, M. and {Asgekar}, A. and {Avruch}, I.~M. and {Batejat}, F. and {B{\"a}hren}, L. and {Bell}, M.~E. and {Bell}, M.~R. and {van Bemmel}, I. and {Bennema}, P. and {Bentum}, M.~J. and {Bernardi}, G. and {Best}, P. and {B{\^\i}rzan}, L. and {Bonafede}, A. and {Boonstra}, A.-J. and {Braun}, R. and {Bregman}, J. and {Breitling}, F. and {van de Brink}, R.~H. and {Broderick}, J. and {Broekema}, P.~C. and {Brouw}, W.~N. and {Br{\"u}ggen}, M. and {Butcher}, H.~R. and {van Cappellen}, W. and {Ciardi}, B. and {Coenen}, T. and {Conway}, J. and {Coolen}, A. and {Corstanje}, A. and {Damstra}, S. and {Davies}, O. and {Deller}, A.~T. and {Dettmar}, R.-J. and {van Diepen}, G. and {Dijkstra}, K. and {Donker}, P. and {Doorduin}, A. and {Dromer}, J. and {Drost}, M. and {van Duin}, A. and {Eisl{\"o}ffel}, J. and {van Enst}, J. and {Ferrari}, C. and {Frieswijk}, W. and {Gankema}, H. and {Garrett}, M.~A. and {de Gasperin}, F. and {Gerbers}, M. and {de Geus}, E. and {Grie{\ss}meier}, J.-M. and {Grit}, T. and {Gruppen}, P. and {Hamaker}, J.~P. and {Hassall}, T. and {Hoeft}, M. and {Holties}, H.~A. and {Horneffer}, A. and {van der Horst}, A. and {van Houwelingen}, A. and {Huijgen}, A. and {Iacobelli}, M. and {Intema}, H. and {Jackson}, N. and {Jelic}, V. and {de Jong}, A. and {Juette}, E. and {Kant}, D. and {Karastergiou}, A. and {Koers}, A. and {Kollen}, H. and {Kondratiev}, V.~I. and {Kooistra}, E. and {Koopman}, Y. and {Koster}, A. and {Kuniyoshi}, M. and {Kramer}, M. and {Kuper}, G. and {Lambropoulos}, P. and {Law}, C. and {van Leeuwen}, J. and {Lemaitre}, J. and {Loose}, M. and {Maat}, P. and {Macario}, G. and {Markoff}, S. and {Masters}, J. and {McFadden}, R.~A. and {McKay-Bukowski}, D. and {Meijering}, H. and {Meulman}, H. and {Mevius}, M. and {Middelberg}, E. and {Millenaar}, R. and {Miller-Jones}, J.~C.~A. and {Mohan}, R.~N. and {Mol}, J.~D. and {Morawietz}, J. and {Morganti}, R. and {Mulcahy}, D.~D. and {Mulder}, E. and {Munk}, H. and {Nieuwenhuis}, L. and {van Nieuwpoort}, R. and {Noordam}, J.~E. and {Norden}, M. and {Noutsos}, A. and {Offringa}, A.~R. and {Olofsson}, H. and {Omar}, A. and {Orr{\'u}}, E. and {Overeem}, R. and {Paas}, H. and {Pandey-Pommier}, M. and {Pandey}, V.~N. and {Pizzo}, R. and {Polatidis}, A. and {Rafferty}, D. and {Rawlings}, S. and {Reich}, W. and {de Reijer}, J.-P. and {Reitsma}, J. and {Renting}, G.~A. and {Riemers}, P. and {Rol}, E. and {Romein}, J.~W. and {Roosjen}, J. and {Ruiter}, M. and {Scaife}, A. and {van der Schaaf}, K. and {Scheers}, B. and {Schellart}, P. and {Schoenmakers}, A. and {Schoonderbeek}, G. and {Serylak}, M. and {Shulevski}, A. and {Sluman}, J. and {Smirnov}, O. and {Sobey}, C. and {Spreeuw}, H. and {Steinmetz}, M. and {Sterks}, C.~G.~M. and {Stiepel}, H.-J. and {Stuurwold}, K. and {Tagger}, M. and {Tang}, Y. and {Tasse}, C. and {Thomas}, I. and {Thoudam}, S. and {Toribio}, M.~C. and {van der Tol}, B. and {Usov}, O. and {van Veelen}, M. and {van der Veen}, A.-J. and {ter Veen}, S. and {Verbiest}, J.~P.~W. and {Vermeulen}, R. and {Vermaas}, N. and {Vocks}, C. and {Vogt}, C. and {de Vos}, M. and {van der Wal}, E. and {van Weeren}, R. and {Weggemans}, H. and {Weltevrede}, P. and {White}, S. and {Wijnholds}, S.~J. and {Wilhelmsson}, T. and {Wucknitz}, O. and {Yatawatta}, S. and {Zarka}, P. and {Zensus}, A.},
        title = "{LOFAR: The LOw-Frequency ARray}",
      journal = {\aap},
         year = 2013,
        month = aug,
       volume = {556},
          eid = {A2},
        pages = {A2},
          doi = {10.1051/0004-6361/201220873},
archivePrefix = {arXiv},
       eprint = {1305.3550},
 primaryClass = {astro-ph.IM},
       adsurl = {https://ui.adsabs.harvard.edu/abs/2013A&A...556A...2V}
}

@ARTICLE{ska-main,
       author = {{Dewdney}, P.~E. and {Hall}, P.~J. and {Schilizzi}, R.~T. and {Lazio}, T.~J.~L.~W.},
        title = "{The Square Kilometre Array}",
      journal = {IEEE Proceedings},
         year = 2009,
        month = aug,
       volume = {97},
       number = {8},
        pages = {1482-1496},
          doi = {10.1109/JPROC.2009.2021005},
       adsurl = {https://ui.adsabs.harvard.edu/abs/2009IEEEP..97.1482D}
}

@ARTICLE{radioastron-megamaser,
       author = {{Baan}, Willem A. and {An}, Tao and {Henkel}, Christian and {Imai}, Hiroshi and {Kostenko}, Vladimir and {Sobolev}, Andrej},
        title = "{H$_{2}$O MegaMaser emission in NGC 4258 indicative of a periodic disc instability}",
      journal = {Nature Astronomy},
         year = 2022,
        month = jun,
       volume = {6},
        pages = {976-983},
          doi = {10.1038/s41550-022-01706-y},
archivePrefix = {arXiv},
       eprint = {2302.14738},
 primaryClass = {astro-ph.GA},
       adsurl = {https://ui.adsabs.harvard.edu/abs/2022NatAs...6..976B}
}

@ARTICLE{radioastron-agn,
       author = {{G{\'o}mez}, Jos{\'e} L. and {Traianou}, Efthalia and {Krichbaum}, Thomas P. and {Lobanov}, Andrei P. and {Fuentes}, Antonio and {Lico}, Rocco and {Zhao}, Guang-Yao and {Bruni}, Gabriele and {Kovalev}, Yuri Y. and {L{\"a}hteenm{\"a}ki}, Anne and {Voitsik}, Petr A. and {Lisakov}, Mikhail M. and {Angelakis}, Emmanouil and {Bach}, Uwe and {Casadio}, Carolina and {Cho}, Ilje and {Dey}, Lankeswar and {Gopakumar}, Achamveedu and {Gurvits}, Leonid I. and {Jorstad}, Svetlana and {Kovalev}, Yuri A. and {Lister}, Matthew L. and {Marscher}, Alan P. and {Myserlis}, Ioannis and {Pushkarev}, Alexander B. and {Ros}, Eduardo and {Savolainen}, Tuomas and {Tornikoski}, Merja and {Valtonen}, Mauri J. and {Zensus}, Anton},
        title = "{Probing the Innermost Regions of AGN Jets and Their Magnetic Fields with RadioAstron. V. Space and Ground Millimeter-VLBI Imaging of OJ 287}",
      journal = {\apj},
         year = 2022,
        month = jan,
       volume = {924},
       number = {2},
          eid = {122},
        pages = {122},
          doi = {10.3847/1538-4357/ac3bcc},
archivePrefix = {arXiv},
       eprint = {2111.11200},
 primaryClass = {astro-ph.HE},
       adsurl = {https://ui.adsabs.harvard.edu/abs/2022ApJ...924..122G}
}

@ARTICLE{eht-imaging,
       author = {{Event Horizon Telescope Collaboration} and {Akiyama}, Kazunori and {Alberdi}, Antxon and {Alef}, Walter and {Asada}, Keiichi and {Azulay}, Rebecca and {Baczko}, Anne-Kathrin and {Ball}, David and {Balokovi{\'c}}, Mislav and {Barrett}, John and {Bintley}, Dan and {Blackburn}, Lindy and {Boland}, Wilfred and {Bouman}, Katherine L. and {Bower}, Geoffrey C. and {Bremer}, Michael and {Brinkerink}, Christiaan D. and {Brissenden}, Roger and {Britzen}, Silke and {Broderick}, Avery E. and {Broguiere}, Dominique and {Bronzwaer}, Thomas and {Byun}, Do-Young and {Carlstrom}, John E. and {Chael}, Andrew and {Chan}, Chi-kwan and {Chatterjee}, Shami and {Chatterjee}, Koushik and {Chen}, Ming-Tang and {Chen}, Yongjun and {Cho}, Ilje and {Christian}, Pierre and {Conway}, John E. and {Cordes}, James M. and {Crew}, Geoffrey B. and {Cui}, Yuzhu and {Davelaar}, Jordy and {De Laurentis}, Mariafelicia and {Deane}, Roger and {Dempsey}, Jessica and {Desvignes}, Gregory and {Dexter}, Jason and {Doeleman}, Sheperd S. and {Eatough}, Ralph P. and {Falcke}, Heino and {Fish}, Vincent L. and {Fomalont}, Ed and {Fraga-Encinas}, Raquel and {Freeman}, William T. and {Friberg}, Per and {Fromm}, Christian M. and {G{\'o}mez}, Jos{\'e} L. and {Galison}, Peter and {Gammie}, Charles F. and {Garc{\'\i}a}, Roberto and {Gentaz}, Olivier and {Georgiev}, Boris and {Goddi}, Ciriaco and {Gold}, Roman and {Gu}, Minfeng and {Gurwell}, Mark and {Hada}, Kazuhiro and {Hecht}, Michael H. and {Hesper}, Ronald and {Ho}, Luis C. and {Ho}, Paul and {Honma}, Mareki and {Huang}, Chih-Wei L. and {Huang}, Lei and {Hughes}, David H. and {Ikeda}, Shiro and {Inoue}, Makoto and {Issaoun}, Sara and {James}, David J. and {Jannuzi}, Buell T. and {Janssen}, Michael and {Jeter}, Britton and {Jiang}, Wu and {Johnson}, Michael D. and {Jorstad}, Svetlana and {Jung}, Taehyun and {Karami}, Mansour and {Karuppusamy}, Ramesh and {Kawashima}, Tomohisa and {Keating}, Garrett K. and {Kettenis}, Mark and {Kim}, Jae-Young and {Kim}, Junhan and {Kim}, Jongsoo and {Kino}, Motoki and {Koay}, Jun Yi and {Koch}, Patrick M. and {Koyama}, Shoko and {Kramer}, Michael and {Kramer}, Carsten and {Krichbaum}, Thomas P. and {Kuo}, Cheng-Yu and {Lauer}, Tod R. and {Lee}, Sang-Sung and {Li}, Yan-Rong and {Li}, Zhiyuan and {Lindqvist}, Michael and {Liu}, Kuo and {Liuzzo}, Elisabetta and {Lo}, Wen-Ping and {Lobanov}, Andrei P. and {Loinard}, Laurent and {Lonsdale}, Colin and {Lu}, Ru-Sen and {MacDonald}, Nicholas R. and {Mao}, Jirong and {Markoff}, Sera and {Marrone}, Daniel P. and {Marscher}, Alan P. and {Mart{\'\i}-Vidal}, Iv{\'a}n and {Matsushita}, Satoki and {Matthews}, Lynn D. and {Medeiros}, Lia and {Menten}, Karl M. and {Mizuno}, Yosuke and {Mizuno}, Izumi and {Moran}, James M. and {Moriyama}, Kotaro and {Moscibrodzka}, Monika and {M{\"u}ller}, Cornelia and {Nagai}, Hiroshi and {Nagar}, Neil M. and {Nakamura}, Masanori and {Narayan}, Ramesh and {Narayanan}, Gopal and {Natarajan}, Iniyan and {Neri}, Roberto and {Ni}, Chunchong and {Noutsos}, Aristeidis and {Okino}, Hiroki and {Olivares}, H{\'e}ctor and {Oyama}, Tomoaki and {{\"O}zel}, Feryal and {Palumbo}, Daniel C.~M. and {Patel}, Nimesh and {Pen}, Ue-Li and {Pesce}, Dominic W. and {Pi{\'e}tu}, Vincent and {Plambeck}, Richard and {PopStefanija}, Aleksandar and {Porth}, Oliver and {Prather}, Ben and {Preciado-L{\'o}pez}, Jorge A. and {Psaltis}, Dimitrios and {Pu}, Hung-Yi and {Ramakrishnan}, Venkatessh and {Rao}, Ramprasad and {Rawlings}, Mark G. and {Raymond}, Alexander W. and {Rezzolla}, Luciano and {Ripperda}, Bart and {Roelofs}, Freek and {Rogers}, Alan and {Ros}, Eduardo and {Rose}, Mel and {Roshanineshat}, Arash and {Rottmann}, Helge and {Roy}, Alan L. and {Ruszczyk}, Chet and {Ryan}, Benjamin R. and {Rygl}, Kazi L.~J. and {S{\'a}nchez}, Salvador and {S{\'a}nchez-Arguelles}, David and {Sasada}, Mahito and {Savolainen}, Tuomas and {Schloerb}, F. Peter and {Schuster}, Karl-Friedrich and {Shao}, Lijing and {Shen}, Zhiqiang and {Small}, Des and {Sohn}, Bong Won and {SooHoo}, Jason and {Tazaki}, Fumie and {Tiede}, Paul and {Tilanus}, Remo P.~J. and {Titus}, Michael and {Toma}, Kenji and {Torne}, Pablo and {Trent}, Tyler and {Trippe}, Sascha and {Tsuda}, Shuichiro and {van Bemmel}, Ilse and {van Langevelde}, Huib Jan and {van Rossum}, Daniel R. and {Wagner}, Jan and {Wardle}, John and {Weintroub}, Jonathan and {Wex}, Norbert and {Wharton}, Robert and {Wielgus}, Maciek and {Wong}, George N. and {Wu}, Qingwen and {Young}, Andr{\'e} and {Young}, Ken and {Younsi}, Ziri},
        title = "{First M87 Event Horizon Telescope Results. IV. Imaging the Central Supermassive Black Hole}",
      journal = {\apjl},
         year = 2019,
        month = apr,
       volume = {875},
       number = {1},
          eid = {L4},
        pages = {L4},
          doi = {10.3847/2041-8213/ab0e85},
archivePrefix = {arXiv},
       eprint = {1906.11241},
 primaryClass = {astro-ph.GA},
       adsurl = {https://ui.adsabs.harvard.edu/abs/2019ApJ...875L...4E}
}

@ARTICLE{alma-super-resolution,
       author = {{Shoshi}, Ayumu and {Yamaguchi}, Masayuki and {Muto}, Takayuki and {Hirano}, Naomi and {Kawabe}, Ryohei and {Tsukagoshi}, Takashi and {Machida}, Masahiro N.},
        title = "{ALMA 2D super-resolution imaging survey of Ophiuchus Class I/flat spectrum/II disks. I. Discovery of new disk substructures}",
      journal = {\pasj},
         year = 2025,
        month = jun,
       volume = {77},
       number = {3},
        pages = {572-606},
          doi = {10.1093/pasj/psaf026},
archivePrefix = {arXiv},
       eprint = {2503.21060},
 primaryClass = {astro-ph.EP},
       adsurl = {https://ui.adsabs.harvard.edu/abs/2025PASJ...77..572S}
}

@ARTICLE{eht-m87-1,
       author = {{Event Horizon Telescope Collaboration} and {Akiyama}, Kazunori and {Alberdi}, Antxon and {Alef}, Walter and {Asada}, Keiichi and {Azulay}, Rebecca and {Baczko}, Anne-Kathrin and {Ball}, David and {Balokovi{\'c}}, Mislav and {Barrett}, John and {Bintley}, Dan and {Blackburn}, Lindy and {Boland}, Wilfred and {Bouman}, Katherine L. and {Bower}, Geoffrey C. and {Bremer}, Michael and {Brinkerink}, Christiaan D. and {Brissenden}, Roger and {Britzen}, Silke and {Broderick}, Avery E. and {Broguiere}, Dominique and {Bronzwaer}, Thomas and {Byun}, Do-Young and {Carlstrom}, John E. and {Chael}, Andrew and {Chan}, Chi-kwan and {Chatterjee}, Shami and {Chatterjee}, Koushik and {Chen}, Ming-Tang and {Chen}, Yongjun and {Cho}, Ilje and {Christian}, Pierre and {Conway}, John E. and {Cordes}, James M. and {Crew}, Geoffrey B. and {Cui}, Yuzhu and {Davelaar}, Jordy and {De Laurentis}, Mariafelicia and {Deane}, Roger and {Dempsey}, Jessica and {Desvignes}, Gregory and {Dexter}, Jason and {Doeleman}, Sheperd S. and {Eatough}, Ralph P. and {Falcke}, Heino and {Fish}, Vincent L. and {Fomalont}, Ed and {Fraga-Encinas}, Raquel and {Freeman}, William T. and {Friberg}, Per and {Fromm}, Christian M. and {G{\'o}mez}, Jos{\'e} L. and {Galison}, Peter and {Gammie}, Charles F. and {Garc{\'\i}a}, Roberto and {Gentaz}, Olivier and {Georgiev}, Boris and {Goddi}, Ciriaco and {Gold}, Roman and {Gu}, Minfeng and {Gurwell}, Mark and {Hada}, Kazuhiro and {Hecht}, Michael H. and {Hesper}, Ronald and {Ho}, Luis C. and {Ho}, Paul and {Honma}, Mareki and {Huang}, Chih-Wei L. and {Huang}, Lei and {Hughes}, David H. and {Ikeda}, Shiro and {Inoue}, Makoto and {Issaoun}, Sara and {James}, David J. and {Jannuzi}, Buell T. and {Janssen}, Michael and {Jeter}, Britton and {Jiang}, Wu and {Johnson}, Michael D. and {Jorstad}, Svetlana and {Jung}, Taehyun and {Karami}, Mansour and {Karuppusamy}, Ramesh and {Kawashima}, Tomohisa and {Keating}, Garrett K. and {Kettenis}, Mark and {Kim}, Jae-Young and {Kim}, Junhan and {Kim}, Jongsoo and {Kino}, Motoki and {Koay}, Jun Yi and {Koch}, Patrick M. and {Koyama}, Shoko and {Kramer}, Michael and {Kramer}, Carsten and {Krichbaum}, Thomas P. and {Kuo}, Cheng-Yu and {Lauer}, Tod R. and {Lee}, Sang-Sung and {Li}, Yan-Rong and {Li}, Zhiyuan and {Lindqvist}, Michael and {Liu}, Kuo and {Liuzzo}, Elisabetta and {Lo}, Wen-Ping and {Lobanov}, Andrei P. and {Loinard}, Laurent and {Lonsdale}, Colin and {Lu}, Ru-Sen and {MacDonald}, Nicholas R. and {Mao}, Jirong and {Markoff}, Sera and {Marrone}, Daniel P. and {Marscher}, Alan P. and {Mart{\'\i}-Vidal}, Iv{\'a}n and {Matsushita}, Satoki and {Matthews}, Lynn D. and {Medeiros}, Lia and {Menten}, Karl M. and {Mizuno}, Yosuke and {Mizuno}, Izumi and {Moran}, James M. and {Moriyama}, Kotaro and {Moscibrodzka}, Monika and {M{\"u}ller}, Cornelia and {Nagai}, Hiroshi and {Nagar}, Neil M. and {Nakamura}, Masanori and {Narayan}, Ramesh and {Narayanan}, Gopal and {Natarajan}, Iniyan and {Neri}, Roberto and {Ni}, Chunchong and {Noutsos}, Aristeidis and {Okino}, Hiroki and {Olivares}, H{\'e}ctor and {Ortiz-Le{\'o}n}, Gisela N. and {Oyama}, Tomoaki and {{\"O}zel}, Feryal and {Palumbo}, Daniel C.~M. and {Patel}, Nimesh and {Pen}, Ue-Li and {Pesce}, Dominic W. and {Pi{\'e}tu}, Vincent and {Plambeck}, Richard and {PopStefanija}, Aleksandar and {Porth}, Oliver and {Prather}, Ben and {Preciado-L{\'o}pez}, Jorge A. and {Psaltis}, Dimitrios and {Pu}, Hung-Yi and {Ramakrishnan}, Venkatessh and {Rao}, Ramprasad and {Rawlings}, Mark G. and {Raymond}, Alexander W. and {Rezzolla}, Luciano and {Ripperda}, Bart and {Roelofs}, Freek and {Rogers}, Alan and {Ros}, Eduardo and {Rose}, Mel and {Roshanineshat}, Arash and {Rottmann}, Helge and {Roy}, Alan L. and {Ruszczyk}, Chet and {Ryan}, Benjamin R. and {Rygl}, Kazi L.~J. and {S{\'a}nchez}, Salvador and {S{\'a}nchez-Arguelles}, David and {Sasada}, Mahito and {Savolainen}, Tuomas and {Schloerb}, F. Peter and {Schuster}, Karl-Friedrich and {Shao}, Lijing and {Shen}, Zhiqiang and {Small}, Des and {Sohn}, Bong Won and {SooHoo}, Jason and {Tazaki}, Fumie and {Tiede}, Paul and {Tilanus}, Remo P.~J. and {Titus}, Michael and {Toma}, Kenji and {Torne}, Pablo and {Trent}, Tyler and {Trippe}, Sascha and {Tsuda}, Shuichiro and {van Bemmel}, Ilse and {van Langevelde}, Huib Jan and {van Rossum}, Daniel R. and {Wagner}, Jan and {Wardle}, John and {Weintroub}, Jonathan and {Wex}, Norbert and {Wharton}, Robert and {Wielgus}, Maciek and {Wong}, George N. and {Wu}, Qingwen and {Young}, Ken and {Young}, Andr{\'e}},
        title = "{First M87 Event Horizon Telescope Results. I. The Shadow of the Supermassive Black Hole}",
      journal = {\apjl},
         year = 2019,
        month = apr,
       volume = {875},
       number = {1},
          eid = {L1},
        pages = {L1},
          doi = {10.3847/2041-8213/ab0ec7},
archivePrefix = {arXiv},
       eprint = {1906.11238},
 primaryClass = {astro-ph.GA},
       adsurl = {https://ui.adsabs.harvard.edu/abs/2019ApJ...875L...1E}
}

@ARTICLE{eht-cagA-1,
       author = {{Event Horizon Telescope Collaboration} and {Akiyama}, Kazunori and {Alberdi}, Antxon and {Alef}, Walter and {Algaba}, Juan Carlos and {Anantua}, Richard and {Asada}, Keiichi and {Azulay}, Rebecca and {Bach}, Uwe and {Baczko}, Anne-Kathrin and {Ball}, David and {Balokovi{\'c}}, Mislav and {Barrett}, John and {Baub{\"o}ck}, Michi and {Benson}, Bradford A. and {Bintley}, Dan and {Blackburn}, Lindy and {Blundell}, Raymond and {Bouman}, Katherine L. and {Bower}, Geoffrey C. and {Boyce}, Hope and {Bremer}, Michael and {Brinkerink}, Christiaan D. and {Brissenden}, Roger and {Britzen}, Silke and {Broderick}, Avery E. and {Broguiere}, Dominique and {Bronzwaer}, Thomas and {Bustamante}, Sandra and {Byun}, Do-Young and {Carlstrom}, John E. and {Ceccobello}, Chiara and {Chael}, Andrew and {Chan}, Chi-kwan and {Chatterjee}, Koushik and {Chatterjee}, Shami and {Chen}, Ming-Tang and {Chen}, Yongjun and {Cheng}, Xiaopeng and {Cho}, Ilje and {Christian}, Pierre and {Conroy}, Nicholas S. and {Conway}, John E. and {Cordes}, James M. and {Crawford}, Thomas M. and {Crew}, Geoffrey B. and {Cruz-Osorio}, Alejandro and {Cui}, Yuzhu and {Davelaar}, Jordy and {De Laurentis}, Mariafelicia and {Deane}, Roger and {Dempsey}, Jessica and {Desvignes}, Gregory and {Dexter}, Jason and {Dhruv}, Vedant and {Doeleman}, Sheperd S. and {Dougal}, Sean and {Dzib}, Sergio A. and {Eatough}, Ralph P. and {Emami}, Razieh and {Falcke}, Heino and {Farah}, Joseph and {Fish}, Vincent L. and {Fomalont}, Ed and {Ford}, H. Alyson and {Fraga-Encinas}, Raquel and {Freeman}, William T. and {Friberg}, Per and {Fromm}, Christian M. and {Fuentes}, Antonio and {Galison}, Peter and {Gammie}, Charles F. and {Garc{\'\i}a}, Roberto and {Gentaz}, Olivier and {Georgiev}, Boris and {Goddi}, Ciriaco and {Gold}, Roman and {G{\'o}mez-Ruiz}, Arturo I. and {G{\'o}mez}, Jos{\'e} L. and {Gu}, Minfeng and {Gurwell}, Mark and {Hada}, Kazuhiro and {Haggard}, Daryl and {Haworth}, Kari and {Hecht}, Michael H. and {Hesper}, Ronald and {Heumann}, Dirk and {Ho}, Luis C. and {Ho}, Paul and {Honma}, Mareki and {Huang}, Chih-Wei L. and {Huang}, Lei and {Hughes}, David H. and {Ikeda}, Shiro and {Impellizzeri}, C.~M. Violette and {Inoue}, Makoto and {Issaoun}, Sara and {James}, David J. and {Jannuzi}, Buell T. and {Janssen}, Michael and {Jeter}, Britton and {Jiang}, Wu and {Jim{\'e}nez-Rosales}, Alejandra and {Johnson}, Michael D. and {Jorstad}, Svetlana and {Joshi}, Abhishek V. and {Jung}, Taehyun and {Karami}, Mansour and {Karuppusamy}, Ramesh and {Kawashima}, Tomohisa and {Keating}, Garrett K. and {Kettenis}, Mark and {Kim}, Dong-Jin and {Kim}, Jae-Young and {Kim}, Jongsoo and {Kim}, Junhan and {Kino}, Motoki and {Koay}, Jun Yi and {Kocherlakota}, Prashant and {Kofuji}, Yutaro and {Koch}, Patrick M. and {Koyama}, Shoko and {Kramer}, Carsten and {Kramer}, Michael and {Krichbaum}, Thomas P. and {Kuo}, Cheng-Yu and {La Bella}, Noemi and {Lauer}, Tod R. and {Lee}, Daeyoung and {Lee}, Sang-Sung and {Leung}, Po Kin and {Levis}, Aviad and {Li}, Zhiyuan and {Lico}, Rocco and {Lindahl}, Greg and {Lindqvist}, Michael and {Lisakov}, Mikhail and {Liu}, Jun and {Liu}, Kuo and {Liuzzo}, Elisabetta and {Lo}, Wen-Ping and {Lobanov}, Andrei P. and {Loinard}, Laurent and {Lonsdale}, Colin J. and {Lu}, Ru-Sen and {Mao}, Jirong and {Marchili}, Nicola and {Markoff}, Sera and {Marrone}, Daniel P. and {Marscher}, Alan P. and {Mart{\'\i}-Vidal}, Iv{\'a}n and {Matsushita}, Satoki and {Matthews}, Lynn D. and {Medeiros}, Lia and {Menten}, Karl M. and {Michalik}, Daniel and {Mizuno}, Izumi and {Mizuno}, Yosuke and {Moran}, James M. and {Moriyama}, Kotaro and {Moscibrodzka}, Monika and {M{\"u}ller}, Cornelia and {Mus}, Alejandro and {Musoke}, Gibwa and {Myserlis}, Ioannis and {Nadolski}, Andrew and {Nagai}, Hiroshi and {Nagar}, Neil M. and {Nakamura}, Masanori and {Narayan}, Ramesh and {Narayanan}, Gopal and {Natarajan}, Iniyan and {Nathanail}, Antonios and {Fuentes}, Santiago Navarro and {Neilsen}, Joey and {Neri}, Roberto and {Ni}, Chunchong and {Noutsos}, Aristeidis and {Nowak}, Michael A. and {Oh}, Junghwan and {Okino}, Hiroki and {Olivares}, H{\'e}ctor and {Ortiz-Le{\'o}n}, Gisela N. and {Oyama}, Tomoaki and {{\"O}zel}, Feryal and {Palumbo}, Daniel C.~M. and {Paraschos}, Georgios Filippos and {Park}, Jongho and {Parsons}, Harriet and {Patel}, Nimesh and {Pen}, Ue-Li and {Pesce}, Dominic W. and {Pi{\'e}tu}, Vincent and {Plambeck}, Richard and {PopStefanija}, Aleksandar and {Porth}, Oliver and {P{\"o}tzl}, Felix M. and {Prather}, Ben and {Preciado-L{\'o}pez}, Jorge A. and {Psaltis}, Dimitrios},
        title = "{First Sagittarius A* Event Horizon Telescope Results. I. The Shadow of the Supermassive Black Hole in the Center of the Milky Way}",
      journal = {\apjl},
         year = 2022,
        month = may,
       volume = {930},
       number = {2},
          eid = {L12},
        pages = {L12},
          doi = {10.3847/2041-8213/ac6674},
       adsurl = {https://ui.adsabs.harvard.edu/abs/2022ApJ...930L..12E}
}

@ARTICLE{lofar-subarc-imaging,
       author = {{Morabito}, L.~K. and {Jackson}, N.~J. and {Mooney}, S. and {Sweijen}, F. and {Badole}, S. and {Kukreti}, P. and {Venkattu}, D. and {Groeneveld}, C. and {Kappes}, A. and {Bonnassieux}, E. and {Drabent}, A. and {Iacobelli}, M. and {Croston}, J.~H. and {Best}, P.~N. and {Bondi}, M. and {Callingham}, J.~R. and {Conway}, J.~E. and {Deller}, A.~T. and {Hardcastle}, M.~J. and {McKean}, J.~P. and {Miley}, G.~K. and {Moldon}, J. and {R{\"o}ttgering}, H.~J.~A. and {Tasse}, C. and {Shimwell}, T.~W. and {van Weeren}, R.~J. and {Anderson}, J.~M. and {Asgekar}, A. and {Avruch}, I.~M. and {van Bemmel}, I.~M. and {Bentum}, M.~J. and {Bonafede}, A. and {Brouw}, W.~N. and {Butcher}, H.~R. and {Ciardi}, B. and {Corstanje}, A. and {Coolen}, A. and {Damstra}, S. and {de Gasperin}, F. and {Duscha}, S. and {Eisl{\"o}ffel}, J. and {Engels}, D. and {Falcke}, H. and {Garrett}, M.~A. and {Griessmeier}, J. and {Gunst}, A.~W. and {van Haarlem}, M.~P. and {Hoeft}, M. and {van der Horst}, A.~J. and {J{\"u}tte}, E. and {Kadler}, M. and {Koopmans}, L.~V.~E. and {Krankowski}, A. and {Mann}, G. and {Nelles}, A. and {Oonk}, J.~B.~R. and {Orru}, E. and {Paas}, H. and {Pandey}, V.~N. and {Pizzo}, R.~F. and {Pandey-Pommier}, M. and {Reich}, W. and {Rothkaehl}, H. and {Ruiter}, M. and {Schwarz}, D.~J. and {Shulevski}, A. and {Soida}, M. and {Tagger}, M. and {Vocks}, C. and {Wijers}, R.~A.~M.~J. and {Wijnholds}, S.~J. and {Wucknitz}, O. and {Zarka}, P. and {Zucca}, P.},
        title = "{Sub-arcsecond imaging with the International LOFAR Telescope. I. Foundational calibration strategy and pipeline}",
      journal = {\aap},
         year = 2022,
        month = feb,
       volume = {658},
          eid = {A1},
        pages = {A1},
          doi = {10.1051/0004-6361/202140649},
archivePrefix = {arXiv},
       eprint = {2108.07283},
 primaryClass = {astro-ph.IM},
       adsurl = {https://ui.adsabs.harvard.edu/abs/2022A&A...658A...1M}
}

@ARTICLE{mightee,
       author = {{Hale}, C.~L. and {Heywood}, I. and {Jarvis}, M.~J. and {Whittam}, I.~H. and {Best}, P.~N. and {An}, Fangxia and {Bowler}, R.~A.~A. and {Harrison}, I. and {Matthews}, A. and {Smith}, D.~J.~B. and {Taylor}, A.~R. and {Vaccari}, M.},
        title = "{MIGHTEE: the continuum survey Data Release 1}",
      journal = {\mnras},
         year = 2025,
        month = jan,
       volume = {536},
       number = {3},
        pages = {2187-2211},
          doi = {10.1093/mnras/stae2528},
archivePrefix = {arXiv},
       eprint = {2411.04958},
 primaryClass = {astro-ph.GA},
       adsurl = {https://ui.adsabs.harvard.edu/abs/2025MNRAS.536.2187H}
}

@ARTICLE{icrf,
       author = {{Charlot}, P. and {Jacobs}, C.~S. and {Gordon}, D. and {Lambert}, S. and {de Witt}, A. and {B{\"o}hm}, J. and {Fey}, A.~L. and {Heinkelmann}, R. and {Skurikhina}, E. and {Titov}, O. and {Arias}, E.~F. and {Bolotin}, S. and {Bourda}, G. and {Ma}, C. and {Malkin}, Z. and {Nothnagel}, A. and {Mayer}, D. and {MacMillan}, D.~S. and {Nilsson}, T. and {Gaume}, R.},
        title = "{The third realization of the International Celestial Reference Frame by very long baseline interferometry}",
      journal = {\aap},
         year = 2020,
        month = dec,
       volume = {644},
          eid = {A159},
        pages = {A159},
          doi = {10.1051/0004-6361/202038368},
archivePrefix = {arXiv},
       eprint = {2010.13625},
 primaryClass = {astro-ph.GA},
       adsurl = {https://ui.adsabs.harvard.edu/abs/2020A&A...644A.159C}
}

@ARTICLE{itrf,
       author = {{Altamimi}, Zuheir and {Rebischung}, Paul and {M{\'e}tivier}, Laurent and {Collilieux}, Xavier},
        title = "{ITRF2014: A new release of the International Terrestrial Reference Frame modeling nonlinear station motions}",
      journal = {Journal of Geophysical Research (Solid Earth)},
         year = 2016,
        month = aug,
       volume = {121},
       number = {8},
        pages = {6109-6131},
          doi = {10.1002/2016JB013098},
       adsurl = {https://ui.adsabs.harvard.edu/abs/2016JGRB..121.6109A}
}

@ARTICLE{ligo-guide,
       author = {{Abbott}, B.~P. and {Abbott}, R. and {Abbott}, T.~D. and {Abraham}, S. and {Acernese}, F. and {Ackley}, K. and {Adams}, C. and {Adya}, V.~B. and {Affeldt}, C. and {Agathos}, M. and {Agatsuma}, K. and {Aggarwal}, N. and {Aguiar}, O.~D. and {Aiello}, L. and {Ain}, A. and {Ajith}, P. and {Alford}, T. and {Allen}, G. and {Allocca}, A. and {Aloy}, M.~A. and {Altin}, P.~A. and {Amato}, A. and {Ananyeva}, A. and {Anderson}, S.~B. and {Anderson}, W.~G. and {Angelova}, S.~V. and {Antier}, S. and {Appert}, S. and {Arai}, K. and {Araya}, M.~C. and {Areeda}, J.~S. and {Ar{\`e}ne}, M. and {Arnaud}, N. and {Arun}, K.~G. and {Ascenzi}, S. and {Ashton}, G. and {Aston}, S.~M. and {Astone}, P. and {Aubin}, F. and {Aufmuth}, P. and {AultONeal}, K. and {Austin}, C. and {Avendano}, V. and {Avila-Alvarez}, A. and {Babak}, S. and {Bacon}, P. and {Badaracco}, F. and {Bader}, M.~K.~M. and {Bae}, S. and {Baker}, P.~T. and {Baldaccini}, F. and {Ballardin}, G. and {Ballmer}, S.~W. and {Banagiri}, S. and {Barayoga}, J.~C. and {Barclay}, S.~E. and {Barish}, B.~C. and {Barker}, D. and {Barkett}, K. and {Barnum}, S. and {Barone}, F. and {Barr}, B. and {Barsotti}, L. and {Barsuglia}, M. and {Barta}, D. and {Bartlett}, J. and {Bartos}, I. and {Bassiri}, R. and {Basti}, A. and {Bawaj}, M. and {Bayley}, J.~C. and {Bazzan}, M. and {B{\'e}csy}, B. and {Bejger}, M. and {Belahcene}, I. and {Bell}, A.~S. and {Beniwal}, D. and {Berger}, B.~K. and {Bergmann}, G. and {Bernuzzi}, S. and {Bero}, J.~J. and {Berry}, C.~P.~L. and {Bersanetti}, D. and {Bertolini}, A. and {Betzwieser}, J. and {Bhandare}, R. and {Bidler}, J. and {Bilenko}, I.~A. and {Bilgili}, S.~A. and {Billingsley}, G. and {Birch}, J. and {Birney}, R. and {Birnholtz}, O. and {Biscans}, S. and {Biscoveanu}, S. and {Bisht}, A. and {Bitossi}, M. and {Bizouard}, M.~A. and {Blackburn}, J.~K. and {Blair}, C.~D. and {Blair}, D.~G. and {Blair}, R.~M. and {Bloemen}, S. and {Bode}, N. and {Boer}, M. and {Boetzel}, Y. and {Bogaert}, G. and {Bondu}, F. and {Bonilla}, E. and {Bonnand}, R. and {Booker}, P. and {Boom}, B.~A. and {Booth}, C.~D. and {Bork}, R. and {Boschi}, V. and {Bose}, S. and {Bossie}, K. and {Bossilkov}, V. and {Bosveld}, J. and {Bouffanais}, Y. and {Bozzi}, A. and {Bradaschia}, C. and {Brady}, P.~R. and {Bramley}, A. and {Branchesi}, M. and {Brau}, J.~E. and {Briant}, T. and {Briggs}, J.~H. and {Brighenti}, F. and {Brillet}, A. and {Brinkmann}, M. and {Brisson}, V. and {Brockill}, P. and {Brooks}, A.~F. and {Brown}, D.~D. and {Brunett}, S. and {Buikema}, A. and {Bulik}, T. and {Bulten}, H.~J. and {Buonanno}, A. and {Buskulic}, D. and {Buy}, C. and {Byer}, R.~L. and {Cabero}, M. and {Cadonati}, L. and {Cagnoli}, G. and {Cahillane}, C. and {Calder{\'o}n Bustillo}, J. and {Callister}, T.~A. and {Calloni}, E. and {Camp}, J.~B. and {Campbell}, W.~A. and {Canepa}, M. and {Cannon}, K.~C. and {Cao}, H. and {Cao}, J. and {Capocasa}, E. and {Carbognani}, F. and {Caride}, S. and {Carney}, M.~F. and {Carullo}, G. and {Casanueva Diaz}, J. and {Casentini}, C. and {Caudill}, S. and {Cavagli{\`a}}, M. and {Cavalier}, F. and {Cavalieri}, R. and {Cella}, G. and {Cerd{\'a}-Dur{\'a}n}, P. and {Cerretani}, G. and {Cesarini}, E. and {Chaibi}, O. and {Chakravarti}, K. and {Chamberlin}, S.~J. and {Chan}, M. and {Chao}, S. and {Charlton}, P. and {Chase}, E.~A. and {Chassande-Mottin}, E. and {Chatterjee}, D. and {Chaturvedi}, M. and {Chatziioannou}, K. and {Cheeseboro}, B.~D. and {Chen}, H.~Y. and {Chen}, X. and {Chen}, Y. and {Cheng}, H.-P. and {Cheong}, C.~K. and {Chia}, H.~Y. and {Chincarini}, A. and {Chiummo}, A. and {Cho}, G. and {Cho}, H.~S. and {Cho}, M. and {Christensen}, N. and {Chu}, Q. and {Chua}, S. and {Chung}, K.~W. and {Chung}, S. and {Ciani}, G.},
        title = "{A guide to LIGO-Virgo detector noise and extraction of transient gravitational-wave signals}",
      journal = {Classical and Quantum Gravity},
         year = 2020,
        month = mar,
       volume = {37},
       number = {5},
          eid = {055002},
        pages = {055002},
          doi = {10.1088/1361-6382/ab685e},
archivePrefix = {arXiv},
       eprint = {1908.11170},
 primaryClass = {gr-qc},
       adsurl = {https://ui.adsabs.harvard.edu/abs/2020CQGra..37e5002A}
}

@ARTICLE{stat-synchrotron,
       author = {{B{\'e}nard}, C. and {Rousseau}, M.},
        title = "{Statistical properties of synchrotron radiation.}",
      journal = {Journal of the Optical Society of America (1917-1983)},
         year = 1974,
        month = jan,
       volume = {64},
        pages = {1433-1444},
          doi = {10.1364/JOSA.64.001433},
       adsurl = {https://ui.adsabs.harvard.edu/abs/1974JOSA...64.1433B}
}

@BOOK{condon-essential-radio-astronomy,
       author = {{Condon}, James J. and {Ransom}, Scott M.},
        title = "{Essential Radio Astronomy}",
         year = 2016,
       adsurl = {https://ui.adsabs.harvard.edu/abs/2016era..book.....C}
}

@ARTICLE{m87-review,
       author = {{Hada}, Kazuhiro and {Asada}, Keiichi and {Nakamura}, Masanori and {Kino}, Motoki},
        title = "{M 87: a cosmic laboratory for deciphering black hole accretion and jet formation}",
      journal = {\aapr},
         year = 2024,
        month = dec,
       volume = {32},
       number = {1},
          eid = {5},
        pages = {5},
          doi = {10.1007/s00159-024-00155-y},
archivePrefix = {arXiv},
       eprint = {2412.07083},
 primaryClass = {astro-ph.HE},
       adsurl = {https://ui.adsabs.harvard.edu/abs/2024A&ARv..32....5H}
}

@ARTICLE{agn-new-targets,
       author = {{Boccardi}, B. and {Ricci}, L. and {Madika}, E. and {Bartolini}, V. and {Bach}, U. and {Grandi}, P. and {Torresi}, E. and {Krichbaum}, T.~P. and {Zensus}, J.~A.},
        title = "{Jet formation studies in AGN: A search for new targets}",
      journal = {\aap},
         year = 2025,
        month = mar,
       volume = {695},
          eid = {A118},
        pages = {A118},
          doi = {10.1051/0004-6361/202453138},
archivePrefix = {arXiv},
       eprint = {2412.19268},
 primaryClass = {astro-ph.HE},
       adsurl = {https://ui.adsabs.harvard.edu/abs/2025A&A...695A.118B}
}

@INPROCEEDINGS{bhex,
       author = {{Johnson}, Michael D. and {Akiyama}, Kazunori and {Baturin}, Rebecca and {Bilyeu}, Bryan and {Blackburn}, Lindy and {Boroson}, Don and {C{\'a}rdenas-Avenda{\~n}o}, Alejandro and {Chael}, Andrew and {Chan}, Chi-kwan and {Chang}, Dominic and {Cheimets}, Peter and {Chou}, Cathy and {Doeleman}, Sheperd S. and {Farah}, Joseph and {Galison}, Peter and {Gamble}, Ronald and {Gammie}, Charles F. and {Gelles}, Zachary and {G{\'o}mez}, Jos{\'e} L. and {Gralla}, Samuel E. and {Grimes}, Paul and {Gurvits}, Leonid I. and {Hadar}, Shahar and {Haworth}, Kari and {Hada}, Kazuhiro and {Hecht}, Michael H. and {Honma}, Mareki and {Houston}, Janice and {Hudson}, Ben and {Issaoun}, Sara and {Jia}, He and {Jorstad}, Svetlana and {Kauffman}, Jens and {Kovalev}, Yuri Y. and {Kurczynski}, Peter and {Lafon}, Robert E. and {Lupsasca}, Alexandru and {Lehmensiek}, Robert and {Ma}, Chung-Pei and {Marrone}, Daniel P. and {Marscher}, Alan P. and {Melnick}, Gary and {Narayan}, Ramesh and {Niinuma}, Kotaro and {Noble}, Scott C. and {Palmer}, Eric J. and {Palumbo}, Daniel C.~M. and {Paritsky}, Lenny and {Peretz}, Eliad and {Pesce}, Dominic and {Plavin}, Alexander and {Quataert}, Eliot and {Rana}, Hannah and {Ricarte}, Angelo and {Roelofs}, Freek and {Shtyrkova}, Katia and {Sinclair}, Laura C. and {Small}, Jeffrey and {Kumara}, Sridharan Tirupati and {Srinivasan}, Ranjani and {Strominger}, Andrew and {Tiede}, Paul and {Tong}, Edward and {Wang}, Jade and {Weintroub}, Jonathan and {Wielgus}, Maciek and {Wong}, George},
        title = "{The Black Hole Explorer: motivation and vision}",
    booktitle = {Space Telescopes and Instrumentation 2024: Optical, Infrared, and Millimeter Wave},
         year = 2024,
       editor = {{Coyle}, Laura E. and {Matsuura}, Shuji and {Perrin}, Marshall D.},
       series = {Society of Photo-Optical Instrumentation Engineers (SPIE) Conference Series},
       volume = {13092},
        month = aug,
          eid = {130922D},
        pages = {130922D},
          doi = {10.1117/12.3019835},
archivePrefix = {arXiv},
       eprint = {2406.12917},
 primaryClass = {astro-ph.IM},
       adsurl = {https://ui.adsabs.harvard.edu/abs/2024SPIE13092E..2DJ}
}

@ARTICLE{ngeht,
       author = {{Doeleman}, Sheperd S. and {Barrett}, John and {Blackburn}, Lindy and {Bouman}, Katherine L. and {Broderick}, Avery E. and {Chaves}, Ryan and {Fish}, Vincent L. and {Fitzpatrick}, Garret and {Freeman}, Mark and {Fuentes}, Antonio and {G{\'o}mez}, Jos{\'e} L. and {Haworth}, Kari and {Houston}, Janice and {Issaoun}, Sara and {Johnson}, Michael D. and {Kettenis}, Mark and {Loinard}, Laurent and {Nagar}, Neil and {Narayanan}, Gopal and {Oppenheimer}, Aaron and {Palumbo}, Daniel C.~M. and {Patel}, Nimesh and {Pesce}, Dominic W. and {Raymond}, Alexander W. and {Roelofs}, Freek and {Srinivasan}, Ranjani and {Tiede}, Paul and {Weintroub}, Jonathan and {Wielgus}, Maciek},
        title = "{Reference Array and Design Consideration for the Next-Generation Event Horizon Telescope}",
      journal = {Galaxies},
         year = 2023,
        month = oct,
       volume = {11},
       number = {5},
          eid = {107},
        pages = {107},
          doi = {10.3390/galaxies11050107},
archivePrefix = {arXiv},
       eprint = {2306.08787},
 primaryClass = {astro-ph.IM},
       adsurl = {https://ui.adsabs.harvard.edu/abs/2023Galax..11..107D}
}

@ARTICLE{boccardi-review,
       author = {{Boccardi}, B. and {Krichbaum}, T.~P. and {Ros}, E. and {Zensus}, J.~A.},
        title = "{Radio observations of active galactic nuclei with mm-VLBI}",
      journal = {\aapr},
         year = 2017,
        month = nov,
       volume = {25},
       number = {1},
          eid = {4},
        pages = {4},
          doi = {10.1007/s00159-017-0105-6},
archivePrefix = {arXiv},
       eprint = {1711.07548},
 primaryClass = {astro-ph.HE},
       adsurl = {https://ui.adsabs.harvard.edu/abs/2017A&ARv..25....4B}
}

@INPROCEEDINGS{astro2020,
       author = {{Doeleman}, Sheperd and {Blackburn}, Lindy and {Dexter}, Jason and {Gomez}, Jose L. and {Johnson}, Michael D. and {Palumbo}, Daniel C. and {Weintroub}, Jonathan and {Farah}, Joseph R. and {Fish}, Vincent and {Loinard}, Laurent and {Lonsdale}, Colin and {Narayanan}, Gopal and {Patel}, Nimesh A. and {Pesce}, Dominic W. and {Raymond}, Alexander and {Tilanus}, Remo and {Wielgus}, Maciek and {Akiyama}, Kazunori and {Bower}, Geoffrey and {Broderick}, Avery and {Deane}, Roger and {Fromm}, Christian Michael and {Gammie}, Charles and {Gold}, Roman and {Janssen}, Michael and {Kawashima}, Tomohisa and {Krichbaum}, Thomas and {Marrone}, Daniel P. and {Matthews}, Lynn D. and {Mizuno}, Yosuke and {Rezzolla}, Luciano and {Roelofs}, Freek and {Ros}, Eduardo and {Savolainen}, Tuomas K. and {Yuan}, Feng and {Zhao}, Guangyao and {Blackburn}, Lindy and {Doeleman}, Sheperd and {Dexter}, Jason and {Gomez}, Jose L. and {Johnson}, Michael D. and {Palumbo}, Daniel C. and {Weintroub}, Jonathan and {Farah}, Joseph R. and {Fish}, Vincent and {Loinard}, Laurent and {Lonsdale}, Colin and {Narayanan}, Gopal and {Patel}, Nimesh A. and {Pesce}, Dominic W. and {Raymond}, Alexander and {Tilanus}, Remo and {Wielgus}, Maciek and {Akiyama}, Kazunori and {Bower}, Geoffrey and {Broderick}, Avery and {Deane}, Roger and {Fromm}, Christian Michael and {Gammie}, Charles and {Gold}, Roman and {Janssen}, Michael and {Kawashima}, Tomohisa and {Krichbaum}, Thomas and {Marrone}, Daniel P. and {Matthews}, Lynn D. and {Mizuno}, Yosuke and {Rezzolla}, Luciano and {Roelofs}, Freek and {Ros}, Eduardo and {Savolainen}, Tuomas K. and {Yuan}, Feng and {Zhao}, Guangyao},
        title = "{Studying Black Holes on Horizon Scales with VLBI Ground Arrays}",
    booktitle = {Bulletin of the American Astronomical Society},
         year = 2019,
       volume = {51},
        month = sep,
          eid = {256},
        pages = {256},
          doi = {10.48550/arXiv.1909.01411},
archivePrefix = {arXiv},
       eprint = {1909.01411},
 primaryClass = {astro-ph.IM},
       adsurl = {https://ui.adsabs.harvard.edu/abs/2019BAAS...51g.256D}
}

@ARTICLE{chassande-mottin,
       author = {{Chassande-Mottin}, {\'E}ric and {Pai}, Archana},
        title = "{Discrete Time and Frequency Wigner─Ville Distribution: Moyal's Formula and Aliasing}",
      journal = {IEEE Signal Processing Letters},
         year = 2005,
        month = jan,
       volume = {12},
       number = {7},
        pages = {508-511},
          doi = {10.1109/LSP.2005.849493},
       adsurl = {https://ui.adsabs.harvard.edu/abs/2005ISPL...12..508C}
}

@article{flandrin-geometry,
title = {Geometry of Affine Time–Frequency Distributions},
journal = {Applied and Computational Harmonic Analysis},
volume = {3},
number = {1},
pages = {10-39},
year = {1996},
issn = {1063-5203},
doi = {https://doi.org/10.1006/acha.1996.0002},
url = {https://www.sciencedirect.com/science/article/pii/S1063520396900020},
author = {Patrick Flandrin and Paulo Goncalvès}
}

@article{isserlis-theorem,
    author = {Isserlis, L.},
    title = {ON A FORMULA FOR THE PRODUCT-MOMENT COEFFICIENT OF ANY ORDER OF A NORMAL FREQUENCY DISTRIBUTION IN ANY NUMBER OF VARIABLES},
    journal = {Biometrika},
    volume = {12},
    number = {1-2},
    pages = {134-139},
    year = {1918},
    month = {11},
    issn = {0006-3444},
    doi = {10.1093/biomet/12.1-2.134},
    url = {https://doi.org/10.1093/biomet/12.1-2.134},
    eprint = {https://academic.oup.com/biomet/article-pdf/12/1-2/134/481266/12-1-2-134.pdf},
}

@MISC{phase-shifter,
       author = {{Mohammed}, A.},
        title = "{Novel methods of digital phase shifting to achieve arbitrary values of time delay}",
         year = 1985,
        month = sep,
       adsurl = {https://ui.adsabs.harvard.edu/abs/1985STIN...8623806M}
}

@book{bendat,
  place={Hoboken, New Jersey},
  edition={4},
  title={Random Data: Analysis and Measurement Procedures},
  publisher="John Wiley \& Sons, Inc.",
  author={Bendat, Julius S. and Piersol, Allan G.},
  year={2010}
}
